# Steady State versus Pulsed Tokamak Reactors

D. J. Segal[1], A. J. Cerfon[2], J. P. Freidberg[1]

1. MIT Plasma Science and Fusion Center, Cambridge MA 02139 USA
2. Courant Institute of Mathematical Sciences, NYU, New York NY 10012

## Abstract

We have carried out a detailed analysis that compares steady state versus pulsed tokamak reactors. The motivations are as follows. Steady state current drive has turned out to be more difficult than expected - it takes too many watts to drive an Ampere, which has a negative effect on power balance and economics. This is partially compensated by the recent development of high temperature REBCO superconductors, which offers the promise of more compact, lower cost tokamak reactors, both steady state and pulsed. Of renewed interest is the reduction in size of pulsed reactors because of the possibility of higher field OH transformers for a given required pulse size. Our main conclusion is that pulsed reactors may indeed be competitive with steady state reactors and this issue should be re-examined with more detailed engineering level studies.

# 1. Introduction

The work presented here re-examines the long held consensus in the US fusion community that a commercial tokamak reactor must operate as a steady state, rather than pulsed, device. There are two basic reasons motivating this re-examination. First, current drive has proven to be more difficult to achieve than originally believed. The efficiency of the most favorable method, lower hybrid current drive, is low in absolute magnitude [1-3]. It just takes too many watts to drive an ampere. This works against steady state reactors.

Second, early analysis of pulsed devices concluded that the need to survive cyclical thermal and mechanical stresses resulted in relatively large, and therefore, economically unattractive reactors [1,4]. Recent advances in technology have the potential to alleviate these problems, and this works in favor of pulsed reactors.

Will these advances be sufficient to make pulsed reactors competitive with steady state reactors? Our analysis attempts to reassess the steady state versus pulsed comparison by including present estimates of current drive efficiency and by taking advantage of the recent development of new high temperature superconductors (HTS) [5-7]. The ideas are as follows. Make the reactor pulsed in order to resolve the current drive problem. Make the OH transformer and toroidal field coils out of HTS, with the possibility of achieving maximum fields of about 23 T. A high toroidal field is expected to improve performance leading to a smaller reactor. Similarly, a high field OH transformer should also reduce the reactor size, since the same flux swing is now possible with a smaller coil radius. Reduced size implies reduced cost. In addition, advanced technologies involving demountable magnet joints and liquid blankets reduce the major component replacement down time. This allows high average power production even with shorter pulses in compact reactors subject to the same number of stress limited cycles compared to larger low field pulsed reactors.

The strategy of our analysis is guided by the basic principle that all relevant design constraints should be identical for both steady state and pulsed reactors. This should, to the maximum extent possible, allow us to make a fair comparison. The analysis begins with definitions of the reactor mission and the primary metric describing reactor desirability. Next, a reasonably comprehensive list of plasma physics constraints is presented, followed by a similar list for nuclear and engineering constraints. As expected, there are many more constraints than degrees of freedom in the design.

A key discussion then selects those constraints that dominate the design of both steady state and pulsed reactors. Many of the constraints overlap but there are crucial differences. Once the constraints driving the designs have been defined, we develop an analytical design model for each type reactor. The models are tested against existing, more sophisticated reactor designs, to demonstrate credibility. With credibility established, we then design a steady state and a pulsed reactor, including sensitivity studies, which enables us to make fair comparisons. These comparisons allows us to draw our conclusions.

A brief summary of these conclusions is given below, and expanded upon in Sec. 8.

- Pulsed reactors are competitive with steady state reactors, and in fact are predicted to be slightly more desirable in terms of several performance measures.
- Our analysis is focused on a 500 MW thermal reactor rather than the usual larger 2500 MW reactors in the literature. Smaller reactors are desirable from an industrial competitiveness point of view. Their designs are driven more by plasma physics than large reactors where technological constraints dominate.
- Both small steady state and pulsed reactors, however, require an enhanced value of the H-mode multiplying factor $H$ above the empirical value $H=1$, in order to achieve power balance. Typically $H \sim 2$ for steady state reactors and $H \sim 1.3$ for pulsed reactors.
- High field is a potential game changer for steady state reactors, improving performance on virtually all fronts.
- High field helps pulsed reactors, but not as much as steady state reactors. The maximum achievable field is advantageous for the OH transformer, but not, however, for the toroidal field (TF). For the TF, there is an optimum, which is below the maximum value achievable technologically.
- Several important problems remain before moving forward with fusion electricity. These include improving $H$, handling the divertor heat load, first wall survival due to neutron wall loading and disruptions, blanket development, robust sustained hollow current density profiles (steady state), development of pulsed HTS magnets for the Ohmic transformer (pulsed), and more accurate analysis of the required pulse length and corresponding cyclical stresses (pulsed).

There is more research to be done and likely new facilities will be needed.

Before proceeding there are several points worth noting. First, the analysis is aimed at a plasma physics audience, rather than fusion engineers. Second, in the spirit of a physics based audience, the analysis presented is virtually entirely analytical. Simplifying approximations are made including the assumption of large aspect ratio. Even so, correct scaling relations are obtained, and calculated values are at least semi-quantitatively accurate. Third, the assumption of large aspect ratio implies that our analysis should not be applied to the spherical tokamak, which tends to be penalized by the use of this approximation.

## 2. Reactor mission and cost metric

As stated in the introduction the research presented here focuses on the design of steady state and pulsed tokamak reactors. The end goal is to make a comparison between these two options to learn whether one or the other is noticeably more attractive from an economic and technological point of view. To carry out the study two high-level definitions are required: (a) the basic reactor mission and (b) a simplified cost metric. The reactor mission and cost metric must be equally applicable to both steady state and pulsed reactors, thus enabling a fair comparison.

Consider first the reactor mission. We define this to be the production of a specified amount of electric output power $P_E$. Many early fusion reactor designs [9-11] were aimed at large $P_E = 1000 \text{ MWe}$ power plants. However, the power situation in the USA today is focused on smaller plants which are faster to build and more flexible in terms of siting plus grid compatibility. To be competitive in today's market, a desirable power plant would deliver about $P_E \approx 250 \text{ MWe}$, a factor of 4 smaller than earlier designs. In keeping with the physics spirit of our study, we can transform from electrical to thermal fusion power produced in the plasma core by assuming a thermal conversion efficiency of $\eta_T = 0.4$ and including the extra power produced by breeding tritium in the blanket. Consequently, we shall take as the basic mission of both reactors a thermal fusion power output given by

$$P_F = 500 \text{ MW} \tag{1}$$

The second quantity of interest is the cost metric. A fusion power plant, like its fission cousin, will be a complex, high tech device. The implication is that the capital cost, as opposed to the operating or fuel costs, will dominate the economics. Furthermore, since revenues are proportional to the net amount of electricity sold, the usual cost metric used to evaluate power plant attractiveness is the capital cost/net electric watt. Detailed fusion reactor designs calculate this critical parameter by summing the costs of each individual component, a lengthy but sound procedure. While the cost/watt is a reliable cost metric, the level of engineering detail required is beyond the scope of the present analysis.

Again, in keeping with the spirit of a physics based study, two simple but plausible measures of capital cost per watt are (a) the toroidal magnetic field energy within the magnet volume per watt and (b) the plasma volume itself per watt. In general both metrics have qualitatively similar behavior as plasma and engineering parameters vary. We shall assume that magnetic field energy per watt $C_{MAG}$ is the primary cost metric. However, also calculated for comparison is the inverse of the volume per watt $P_{VOL}$, which is the more familiar plasma power density. Either metric leads to great simplifications in the analysis, as well as providing physical intuition as to how improved plasma physics performance can reduce cost. Based on this discussion we define the primary and secondary cost metrics for our studies as

$$\begin{aligned} C_{MAG} &= \frac{W_{TF}}{P_F} \quad \text{MJ/MW} \quad &\text{Primary cost metric} \\ P_{VOL} &= \frac{P_F}{V_P} \quad \text{MW/m}^3 \quad &\text{Secondary inverse cost metric} \end{aligned} \tag{2}$$

where $W_{TF}$ is the toroidal magnetic field energy in the plasma and $V_P$ is the plasma volume.

A comparison of the values of $C_{MAG}$ for steady state and pulsed reactors producing the same thermal fusion power will be the basis for deciding the relative attractiveness of each option.

## 3. Design strategy

The mission of both the steady state and pulsed reactors is to produce a desired fusion power $P_F$ as economically as possible, subject to a large number of physics, nuclear, and engineering constraints. To "design" each reactor we must determine values for the following basic variables,

$$
\begin{array}{ll}
B_0 & \text{Central plasma magnetic field (T)} \\
R_0 & \text{Major radius (m)} \\
I_M & \text{Toroidal plasma current (MA)} \\
\overline{n}_{20} & \text{Average electron density } (10^{20} \text{ m}^{-3}) \\
\overline{T}_k & \text{Average temperature (keV)} \\
P_A & \text{Absorbed RF power (MW)}
\end{array}
\tag{3}
$$

Here, variables with an overbar are volume averaged quantities.

In general, there are more constraints than degrees of freedom in the design. The challenge is to identify the most stringent set of constraints for each reactor type. The designs can then be carried out and the other, unutilized constraints can be tested a posteriori to show that they are satisfied. The identification of the most stringent constraints is discussed in Sections 6 and 7.

Most, but not all, of the constraints driving the designs overlap for both the steady state and pulsed options. The one major difference is related to the method of producing the plasma current. This is discussed in detail as the analysis progresses.

Ultimately, application of the various constraints allows us to express each of the design variables in terms of $B_C$, the maximum field on the inside of the toroidal field magnet. Clearly, $B_C \leq B_{max}$ where $B_{max} \approx 23$ T is the maximum practically achievable magnetic field for the recently developed REBCO high temperature superconductors [3] as applied to toroidal field (TF) magnets and Ohmic (OH) transformers. Of particular interest are expressions for the major radius $R_0 = R_0(B_C)$, and the cost metric $C_{MAG} = C_{MAG}(B_C)$. These are exactly the quantities needed to determine whether or not high field superconductors are a potential game changer for fusion reactor attractiveness. The analysis leads to an optimum value of $B_C$ for each option, which is then substituted to determine the final designs.

It is also worth noting that the scaling relations with $B_C$ contain coefficients that are functions of the various constraints that have been applied. For instance, several key plasma physics parameters related to the constraints involve the elongation $\kappa$, inverse aspect ratio $\varepsilon$, Greenwald density fraction $N_G$, beta normal $\beta_N$, kink safety factor $q_*$, confinement enhancement factor $H$, and current drive efficiency $\eta_{CD}$. Similar engineering and nuclear parameters also appear. The dependencies on these parameters are maintained analytically, thereby making it straightforward to determine which constraints lead to the largest sensitivities in the designs.

While the strategy described above makes logical sense, it is critical to acknowledge at the outset that a fundamental problem arises when carrying out the analysis. Specifically, in none of our designs, nor in fact in any of those presented in the literature, does the final reactor satisfy the criteria of "scientific and engineering credibility" using only the standard, well established values for the constraint limits. Using standard limits invariably leads to reactors with either too low a fusion gain or too large a major radius. To obtain an attractive design some form of enhanced plasma performance is required, usually enhanced confinement (i.e. higher $H > 1$) [3,12,13] or sometimes an enhanced density (i.e. higher $N_G > 1$) [12,13].

For the present analysis we face up to this fundamental problem as follows. We eliminate $P_A$, which is inversely proportional to the fusion gain $Q$, as one of the basic design variables, and replace it with the confinement enhancement factor $H$. The desired fusion gain is then treated as an input, and the design then yields the required value of $H$ for a credible reactor. This value is then compared with the standard value, $H = 1$, to see how much enhancement is needed.

To summarize, the basic unknown design variables given in Eq. (3) are replaced by

$$
\begin{array}{ll}
B_0 & \text{Central plasma magnetic field (T)} \\
R_0 & \text{Major radius (m)} \\
I_M & \text{Toroidal plasma current (MA)} \\
\bar{n}_{20} & \text{Average electron density } (10^{20} \text{ m}^{-3}) \\
\bar{T}_k & \text{Average temperature (keV)} \\
H & \text{Confinement enhancement factor}
\end{array}
\quad (4)
$$

with $Q \propto 1/P_A$ now one of the inputs. This is the reformulated strategy. The next step is to list the various physics, nuclear, and engineering constraints that enter the reactor designs.

## 4. Plasma physics constraints

As stated, there are more constraints than degrees of freedom in the design. In Section 4, we simply state the plasma physics constraints but do not single out the strictest ones that dominate the design of either steady state or pulsed reactors. These choices are made in Sections 6 and 7 where we carry out the design analysis.

### 4.1 Tokamak geometric model and profiles

The analysis begins by illustrating the simple geometric model of our large aspect ratio tokamak reactor. See Fig. 1. The model will be used for both steady state and pulsed reactors. Observe that the plasma cross section is assumed to be elliptical in shape. The toroidal field (TF) coils and blanket have a rectangular cross section while the OH transformer is a circular solenoid with a height equal to that of the TF coils. The choices for the geometric shapes are deliberately chosen to be simple to avoid a false impression of more accuracy than is justified. Even so, the shapes do not alter the basic scaling relations that arise and lead to values for the quantities of interest that are semi-quantitatively accurate.

In addition to the geometric model, several of the constraints require averages over density, temperature, and current density profiles. To carry out these averages we introduce a radial-like normalized flux label $\rho = \rho(\psi)$, with $\psi$ the poloidal magnetic flux, such that $0 \leq \rho \leq 1$ with $\rho = 1$ corresponding to the 95% flux surface. A key point is that for analytic simplicity we approximate the flux surfaces as concentric ellipses defined by

$$\left.\begin{array}{l} R = R_0 + a\rho \cos\alpha \\ Z = \kappa a\rho \sin\alpha \end{array}\right\} \quad \rightarrow \quad \frac{(R-R_0)^2}{a^2} + \frac{Z^2}{\kappa^2 a^2} = \rho^2(\psi) \tag{5}$$

with $\kappa$ the elongation of each flux surface and $0 \leq \alpha \leq 2\pi$. This approximation is obviously not self-consistent with the Grad-Shafranov equation, which provides an exact description of MHD equilibria of tokamaks [14,15]. Even so, it suffices for present purposes where the only need is to evaluate global volume and area integrals, thus capturing the main effects of elongation. Using this approximation, and considering the large aspect ratio limit, it then follows that volume and cross sectional area integrals, within a given flux surface, are related by

$$\int G(R,Z)\, d\mathbf{r} \approx 2\pi R_0 \int G(R,Z)\, dA = 2\pi R_0 a^2 \kappa \int_0^\rho \int_0^{2\pi} G(\rho,\alpha) \rho\, d\rho\, d\alpha \tag{6}$$

Based on this discussion, we choose simple monotonic profiles for the (electron) density $n_e \equiv n$ and temperature profiles $T_e = T_i \equiv T$, modeling those observed in high performance H-mode discharges [16,17]. These are given by

$$\begin{aligned} n(\rho) &= \bar{n}(1+\nu_n)(1-\rho^2)^{\nu_n} \\ T(\rho) &= \bar{T}(1+\nu_T)(1-\rho^2)^{\nu_T} \end{aligned} \tag{7}$$

The quantities $\nu_n$ and $\nu_T$ are profile parameters. During H-mode operation, the density is relatively flat while the temperature is peaked. For numerical substitutions we choose $\nu_n = 0.4$ and $\nu_T = 1.1$.

The poloidally averaged current density is more complicated since its profile will likely be peaked off axis for steady state reactors and peaked on axis for pulsed reactors. A single model that allows for both of these situations as well as being simple enough to carry out certain integrals analytically has the (unintuitive) form

$$\bar{J}_\phi(\rho) \equiv \frac{1}{2\pi} \int_0^{2\pi} J_\phi(\rho,\alpha)\,d\alpha = -\frac{I}{\pi a^2 \kappa}\left\{ 4(1-\nu_J)^2 \frac{(1-\rho^2)(1+3\nu_J \rho^2)}{[1+(1-3\nu_J)\rho^2+\nu_J \rho^4]^3} \right\} \tag{8}$$

where $\nu_J$ is a profile parameter. Physical solutions exist for $-1/3 < \nu_J < 1$ with off axis peaked profiles corresponding to $\nu_J > 1/3$. Note that $\nu_J = -1/3$ corresponds to the practical limit of no current reversal on the inboard midplane while $\nu_J = 1$ produces infinite currents on the plasma surface. Examples of the profiles are illustrated in Fig. 2. The specific choices for $\nu_J$ are discussed in more detail during the calculation of the bootstrap current.

### 4.2 Plasma physics constraints and derived quantities

With the background just provided, we now proceed to list the various plasma physics constraints that must be satisfied by both steady state and pulsed reactors. Also listed are important derived quantities that are required for the analysis.

- **Elongation constraint**

The elongation of the minor cross section of the plasma, denoted by $\kappa$, is limited by resistive wall vertical instabilities. If $\kappa$ is too large a major disruption occurs. In practice, the maximum achievable elongation is set by the properties of the vertical position feedback control system and the ratio of wall to plasma radii. Both theory and experiment indicate that for typical aspect ratios and plausible feedback systems the maximum elongation corresponding robust, reliable operation, is limited by [3,18-20]

$$\kappa \leq 1.8 \tag{9}$$

where $\kappa$ refers to the 95% flux surface.

- **Aspect ratio constraint**

There is no specific plasma instability that sets a limit on the allowable inverse aspect ratio $\varepsilon = a/R_0$. Still, we shall define a restricted range of allowable inverse aspect ratios based on several observations. First is the recognition that the tokamak data base [21] determining the confinement time $\tau_E$ has been obtained over a relatively narrow range of $\varepsilon$. Straying far from this range leads to "fear of the unknown" – that is, a higher risk of uncertainty in predicting $\tau_E$. Second, we see intuitively that a thin bicycle tire plasma leads to a high power reactor[1] while a small holed doughnut presents difficulties fitting all the coils and blankets within the central hole. Based on this reasoning we restrict the aspect ratio to lie in the range

$$2.5 \leq R_0/a \leq 4 \quad \rightarrow \quad 0.25 \leq \varepsilon \leq 0.4 \qquad (10)$$

Clearly, once the inverse aspect ratio is specified, we see that the minor radius and major radius are related by $a = \varepsilon R_0$.

- **Greenwald density constraint**

The well-known Greenwald density limit [22] in practical units is given by

$$\bar{n}_{20} = N_G \frac{I_M}{\pi a^2} = K_G \frac{I_M}{R_0^2} \qquad K_G = 0.3183 \left( \frac{N_G}{\varepsilon^2} \right) \qquad (11)$$

where $\bar{n}_{20} = \bar{n}_e (10^{20} \, \text{m}^{-3})$. The actual density limit appearing in the literature is specified in terms of the line averaged rather than volume averaged density. However, for reasonably flat H-mode density profiles these two averages are nearly equal. Consequently, for mathematical simplicity we shall use the volume average density. The coefficient $N_G$ represents the

---

[1] The reason is that for economic optimization the plasma minor radius must always be greater than or comparable to the blanket thickness, which is of order 1 m. Therefore a bicycle tire has a fixed minimum size minor radius but an increasing major radius, thereby leading to large output powers

Greenwald density fraction and experimental data indicates that $N_G \leq N_{max} = 1$ to avoid major disruptions.

There is also another limit that requires the density to be above a certain value in order to gain access to H-mode operation. This can be seen from the tokamak transport data base [21], as illustrated in Fig. 3. Shown here is the number of discharges that enter H-mode operation as a function of the Greenwald density fraction. We see that $N_G \geq N_{min} = 0.3$ for H-mode access.

The conclusion is that the Greenwald density limit is given by Eq. (11) with $N_G$ constrained to lie in the range

$$N_{min} \leq N_G \leq N_{max} \quad \rightarrow \quad 0.3 \leq N_G \leq 1 \tag{12}$$

In the actual designs, a safety margin is added at each end limit.

- **The Troyon beta constraint**

The Troyon beta limit, based on extensive MHD computational studies and experimental data, is given by [23]

$$\beta \leq \beta_N \frac{I_M}{aB_0} \tag{13}$$

with $\beta_N \leq 0.028 = 2.8\%$ and the surrounding wall assumed to be at infinity. This criterion must be satisfied to avoid MHD enhanced transport or major disruptions.

The actual situation is somewhat more complicated. If the plasma is surrounded by a close fitting, highly conducting, resistive wall, then a combination of feedback and plasma rotation can raise the value of $\beta_N$ towards the perfectly conducting wall limit [24-26] which is substantially higher than the no-wall limit: $\beta_N$ increases to about 4% - 6%. This has been observed experimentally [27,28] but is still not sufficiently robust to be considered "standard" high performance operation. It is an area where research should be continued. For present purposes, we impose the original Troyon value $\beta_N = 0.028$ as the beta limit.

The definition of beta is $\beta = 2\mu_0 \bar{p} / B_0^2$ where $\bar{p} = \bar{p}_e + \bar{p}_D + \bar{p}_T + \bar{p}_\alpha + \bar{p}_Z$ includes the electron, deuterium, tritium, alpha particle, and impurity pressures. For the relatively high densities anticipated in a tokamak reactor, where the electron and ion temperatures are equilibrated, we make the simple approximation that $p \approx 2n_e T = 2nT$. For our profiles this implies that

$$\bar{p} = 2\frac{(1+\nu_n)(1+\nu_T)}{(1+\nu_n+\nu_T)}\bar{n}\bar{T} \tag{14}$$

Equation (14) allows us to rewrite Eq. (13) in practical units as

$$\bar{n}_{20}\bar{T}_k \leq K_\beta \frac{I_M B_0}{R_0} \qquad K_\beta = 12.42 \frac{(1+\nu_n+\nu_T)}{(1+\nu_n)(1+\nu_T)}\left(\frac{\beta_N}{\varepsilon}\right) \tag{15}$$

- **Kink safety factor constraint**

The kink safety factor sets a limit on the maximum ratio $I / B_0$ that must be satisfied in order to avoid major disruptions. The stability limit is usually expressed as

$$\frac{2\pi B_0 a^2}{\mu_0 R_0 I}\left[\frac{1+\kappa^2}{2}\right] \geq q_* \tag{16}$$

Here, the critical kink safety factor, as determined by both computation and experiment [29], is in the range $q_* \approx 1.5 - 2$, depending on the current density profile. Converting to practical units leads to

$$\frac{I_M}{B_0 R_0} \leq K_q \qquad K_q = \frac{5\varepsilon^2 \kappa^{1.27}}{q_*} \tag{17}$$

where we have approximated $(1+\kappa^2)/2 \approx \kappa^{1.27}$ over the range $1 < \kappa < 2$. This leads to an error of at most 10% over this range or 6% for $\kappa \leq 1.8$.

- **Fusion power mission constraint**

The fusion power is obtained from the familiar reaction rate expression

$$P_F = \int E_F n_D n_T \langle \sigma v \rangle d\mathbf{r} \qquad (18)$$

Here, the fusion energy per reaction is $E_F = 17.6 \text{ MeV}$. This energy is apportioned between the neutrons and alpha particles as follows: $E_n = 14.1 \text{ MeV}$, $E_\alpha = 3.5 \text{ MeV}$. For a 50-50 D-T mixture the deuterium and tritium densities are related to the electron density $n_e$ by $n_D = n_T = f_D n_e / 2$ where $f_D$ is the fuel dilution factor due to the presence of alpha particles and impurities. We shall assume that $f_D = 0.85$ for numerical substitution but maintain it symbolically in the analysis to test sensitivity.

To evaluate $P_F$ we make use of Eq. (6) leading to a relatively simple expression given by

$$P_F = \pi^2 E_F f_D^2 R_0 a^2 \kappa \int_0^1 n_e^2 \langle \sigma v \rangle \rho\, d\rho \qquad (19)$$

In practical units Eq. (19) reduces to

$$\begin{aligned} P_F &= \left[ 278.3 f_D^2 \varepsilon^2 \kappa \right] \bar{n}_{20}^2 R_0^3 (\sigma v) \quad \text{MW} \\ (\sigma v) &= 10^{21}(1+\nu_n)^2 \int_0^1 (1-\rho^2)^{2\nu_n} \langle \sigma v \rangle \rho\, d\rho \end{aligned} \qquad (20)$$

The normalized $(\sigma v)$ integral (with curved parentheses) can in principle be easily evaluated numerically using the standard Bosch and Hale analytic form of $\langle \sigma v \rangle$ [30]. However, for purposes of analytic simplicity, we instead use the well-known quadratic approximation for $\langle \sigma v \rangle$, which is reasonably accurate in the regime of interest (i.e. $7 < \bar{T}_k < 25$),

$$\langle \sigma v \rangle \approx 10^{-24} T_k^2 \quad \text{m}^3/\text{sec} \qquad (21)$$

The relative error $1 - \langle \sigma v \rangle_{Anal} / \langle \sigma v \rangle_{Bosch}$ is less than 20% over the range of interest. Integrating over profiles then leads to

$$(\sigma v) = 5 \times 10^{-4} \frac{(1+\nu_n)^2 (1+\nu_T)^2}{1 + 2\nu_n + 2\nu_T} \overline{T}_k^2 \qquad (22)$$

and

$$P_F = K_F \overline{n}_{20}^2 \overline{T}_k^2 R_0^3 \quad \text{MW} \qquad K_F = 0.1392 \frac{(1+\nu_n)^2 (1+\nu_T)^2}{1 + 2\nu_n + 2\nu_T} f_D^2 \varepsilon^2 \kappa \qquad (23)$$

It also follows from the D-T reaction that the alpha and neutron powers are given by

$$\begin{aligned} P_\alpha &= \frac{1}{5} P_F \\ P_n &= \frac{4}{5} P_F \end{aligned} \qquad (24)$$

- **Thermal conduction loss**

The thermal conduction loss $P_\kappa$ is expressed in terms of the energy confinement time of the thermal plasma (i.e. electrons, deuterons, tritons) in the standard way. Specifically, we make use of the fact that for equal temperatures the internal energy of the thermal particles has the form $U = (3/2)(n_e + n_D + n_T)T = (3/2)(1 + f_D)nT$

$$\begin{aligned} P_\kappa &= \frac{1}{\tau_E} \int U d\mathbf{r} = \frac{6\pi^2}{\tau_E} (1+\nu_n)(1+\nu_T)(1+f_D) R_0 a^2 \kappa \overline{n} \overline{T} \int_0^1 (1-\rho^2)^{\nu_n + \nu_T} \rho \, d\rho \\ &= K_\kappa \frac{\overline{n}_{20} \overline{T}_k R_0^3}{\tau_E} \quad \text{MW} \qquad K_\kappa = 0.4744 \frac{(1+\nu_n)(1+\nu_T)}{(1+\nu_n+\nu_T)} (1+f_D) \varepsilon^2 \kappa \end{aligned} \qquad (25)$$

In this expression, we assume that the energy confinement time $\tau_E$ corresponds to operation in the ELMy H-mode regime. This is a reasonably favorable regime and the extensive tokamak database indicates that the empirical energy confinement time, $\tau_E = \tau^{IBP98(y,2)}$ is given by [21]

$$\tau_E = 0.145 H \frac{I_M^{0.93} R_0^{1.39} a^{0.58} \kappa^{0.78} \bar{n}_{20}^{-0.41} B_0^{0.15} A^{0.19}}{(P_\alpha + P_A)^{0.69}} \quad \text{sec} \tag{26}$$

The undefined quantities are (a) $P_A(\text{MW})$, the absorbed auxiliary RF power, (b) $A = 2.5$, the mass number for a 50-50 D-T fuel mixture, and (c) $H$, the H-mode enhancement factor. By definition, the database requires that $H = 1$ although there is a non-negligible spread in the data. The $H$ factor is maintained as one of the basic design variables.

- **Fusion gain constraint**

A steady state fusion reactor requires a certain amount of auxiliary power to maintain the plasma during normal operation. As stated above, the power actually absorbed by the plasma is denoted by $P_A$. For a steady state reactor, the main function of the auxiliary power is to drive a fraction of the toroidal current. It also simultaneously heats the plasma. For present purposes it is assumed that current drive is generated primarily by lower hybrid waves (LHCD). This is the most efficient RF method for driving current.

For a pulsed reactor, the auxiliary heating is assumed to be provided by ion cyclotron heating (ICH). Actually, during flat top operation, no auxiliary power is hypothetically required – the plasma can in principle operate in a fully ignited mode. Still, to make a fair comparison and allow some measure of profile control, we assume an amount of flat-top ICH power is provided that achieves the same recirculating power fraction as required in the steady state system. This assumption makes little difference in the final pulsed design.

For a steady state reactor, the auxiliary power cannot be too high or else the recirculating power fraction $f_{RP}$ of the plant becomes unacceptably large from an economic point of view. We shall assume a maximum allowable value for $f_{RP} \approx 0.15$. This value can be related to the fusion gain $Q = P_F / P_A$, which is the parameter usually calculated in plasma physics. The relationship is obtained from simple power balance,

$$f_{RP} = \frac{P_{RF}}{P_E} = \frac{P_A / \eta_{RF}}{(1 + E_L / E_F)\eta_T P_F} = \frac{1}{(1 + E_L / E_F)\eta_T \eta_{RF} Q}$$

$$P_{RF} = \frac{P_A}{\eta_{RF}} = \frac{P_F}{\eta_{RF} Q} \qquad (27)$$

$$P_E = \left(1 + \frac{E_L}{E_F}\right)\eta_T P_F$$

Here, $P_{RF}$ is the wall supplied RF power, $\eta_{RF}$ is the wall to plasma absorption efficiency, $\eta_T \approx 0.4$ is the thermal conversion efficiency, $E_L / E_F = 4.8 / 17.6 = 0.273$ represents an additional gain in thermal energy produced by breeding tritium in the blanket, and $P_E = (1 + E_L / E_F)\eta_T P_F \approx 255$ MW For LHCD $\eta_{RF} \approx 0.5$ while for ICH $\eta_{RF} \approx 0.75$. Substituting these values, assuming $f_{RP} = 0.15$, we find that the fusion gain constraint reduces to

$$\begin{aligned} P_A &= \frac{P_F}{Q} \qquad Q > 26 \qquad \text{Steady state reactor} \\ P_A &= \frac{P_F}{Q} \qquad Q > 17 \qquad \text{Pulsed reactor} \end{aligned} \qquad (28)$$

In practice, the fusion gain constraint plays an important role in the design of a steady state reactor because of low current drive efficiency. That is, the amount of LHCD power corresponding to $Q = 26$ is not enough to drive the required portion of the steady state current unless the confinement time is enhanced (i.e. $H > 1$). For a pulsed reactor the $Q$ constraint does not play such a major role since no current drive is required.

- **The L-H transition constraint**

It is well known [31] that sufficient total heating power (i.e. alpha plus auxiliary power) must be supplied to the plasma for energy transport to transition from the unfavorable L-mode regime to the more favorable H-mode regime. The value of this transition power has been determined empirically from experimental observations [32,33] and is given by

$$P_{LH} = 0.0488\, \bar{n}_{20}^{0.717} B_0^{0.803} S^{0.941} \left(\frac{2}{A}\right)^{1.1} Z_{main} \approx \left[1.21\, \varepsilon^{0.94} \kappa^{0.56}\right] \bar{n}_{20}^{0.72} B_0^{0.80} R_0^{1.88} \quad \text{MW} \quad (29)$$

Here, $A = 2.5$ is the average mass number of the D-T fuel, $Z_{main} = 1$ is the charge number of the dominant ions (D,T) and $S$ is the plasma surface area, which for an ellipse reduces to

$$S = 8\pi R_0 a \kappa E(k) \approx 39.48 \varepsilon \kappa^{0.6} R_0^2 \quad (30)$$

with $E(k)$ the complete elliptic integral of the second kind and $k^2 = (\kappa^2 - 1)/\kappa^2$. In the regime $1 < \kappa < 2$ we approximate $\kappa E \approx (\pi/2)\kappa^{0.6}$. The transition constraint on the heating power thus reduces to

$$P_\alpha + P_A \geq P_{LH} = K_{LH} \bar{n}_{20}^{0.72} B_0^{0.80} R_0^{1.88} \quad \text{MW} \qquad K_{LH} = 1.21\, \varepsilon^{0.94} \kappa^{0.56} \quad (31)$$

The value of $P_\alpha$ is usually sufficiently large to satisfy the constraint for both steady state and flat-top operation of pulsed devices. However, during start-up the situation is more difficult since alpha power will not be present. We anticipate that a sophisticated time evolution of the density and perhaps the current may be needed to first enter H-mode during start-up. Some additional RF power supplies may also be needed which will add to the capital cost, but will not affect $Q$ since they can be turned off during steady state or flat-top operation. For simplicity, we do not consider the start-up constraints in our analysis.

- **Bootstrap current fraction**

The total toroidal current $I$ flowing in the reactor is comprised of two components. For a steady state reactor these are the bootstrap current $I_B$ and the externally driven RF current $I_{CD}$. For a pulsed reactor they are the bootstrap current $I_B$ and the transformer induced current $I_\Omega$. We can express these combinations mathematically as follows

$$\begin{aligned} I_B + I_{CD} = I &\rightarrow f_B + f_{CD} = 1 \quad \text{Steady state} \\ I_B + I_\Omega = I &\rightarrow f_B + f_\Omega = 1 \quad \text{Pulsed} \end{aligned} \qquad (32)$$

Here, $f_B$, $f_{CD}$, $f_\Omega$ are the corresponding fractional contributions.

In this subsection we focus on the bootstrap current fraction $f_B$, which is an important parameter that enters in the design of both steady state and pulsed reactors. Its value is important in order to determine how much additional current must be provided. Obtaining reasonable accuracy requires a substantial amount of analysis, which is presented in Appendix A. The results are summarized below.

The analysis is based on an expression for the bootstrap current valid for arbitrary cross section assuming (1) equal temperature electrons and ions $T_e = T_i = T$, (2) large aspect ratio $\varepsilon \ll 1$, and (3) negligible collisionality $\nu_* \to 0$ [34]. Under these assumptions the bootstrap current $\mathbf{J}_B \approx J_B \mathbf{e}_\phi$ has the form

$$J_B(\rho) = -3.32 f_T R_0 n T \left[ \frac{1}{n}\frac{dn}{d\psi} + 0.054 \frac{1}{T}\frac{dT}{d\psi} \right] \qquad (33)$$

Here, $f_T(\psi) \approx 1.46 \varepsilon^{1/2} \rho^{1/2}$ is an approximate expression for the trapped particle fraction.

The analysis in Appendix A shows that Eq. (33), using the profiles in Eqs. (7) and (8) plus the elliptic flux surface assumption, leads to an expression for the bootstrap fraction that can be written as

$$\begin{aligned} f_B &= \frac{I_B}{I} = \frac{2\pi a^2 \kappa}{I} \int_0^1 J_B \rho\, d\rho = K_b \frac{\bar{n}_{20} \bar{T}_k R_0^2}{I_M^2} \\ K_b &= 0.6099\, \varepsilon^{5/2} \kappa^{1.27}(1+\nu_n)(1+\nu_T)(\nu_n + .054\nu_T)C_B \end{aligned} \qquad (34)$$

where the coefficient $C_B(\nu_J, \nu_p)$, with $\nu_p = \nu_n + \nu_T$, has the form

$$C_B(\nu_J, \nu_p) = \frac{1}{(1-\nu_J)^2} \int_0^1 x^{1/4}(1-x)^{\nu_p - 1}[1 + (1-3\nu_J)x + \nu_J x^2]^2 dx \tag{35}$$

A complicated but analytic expression for $C_B(\nu_J, \nu_p)$ is given in Appendix A.

Now, to determine the value of $\nu_J$, recall that this is a profile parameter characterizing the shape of the total area averaged $\bar{J}_\phi(\rho)$. For steady state reactors, the value of $\nu_J$ is determined by assuming that current drive is provided primarily by lower hybrid waves (LHCD). These waves produce a LHCD current density profile with an off-axis peak whose location is designed to approximately overlap with that of the bootstrap current. This constraint is discussed in more detail in Appendix A and leads to an approximate form for $\nu_J$ given by

$$\nu_J \approx 0.453 - 0.1(\nu_p - 1.5) \tag{36}$$

For pulsed reactors, which have relatively high current, the value of $\nu_J$ is determined by simultaneously satisfying two constraints: (a) $q_0 \approx 1$ corresponding to the expected sawtooth operation, and (b) $q_* > 2.5$ to avoid current driven disruptions. Appendix A shows that the resulting value of $\nu_J$ has the value

$$\nu_J = 1 - \left(\frac{q_*}{4q_0}\right)^{1/2} = 0.209 \tag{37}$$

with the numerical value corresponding to $q_* = 2.5$. Pulsed reactors have a total current density profile that is peaked on axis.

- **Current drive constraint**

In a steady state reactor, the current drive constraint is one of the dominant drivers of the design. We shall assume that current drive (CD) is provided primarily by lower hybrid waves because of the corresponding relatively high efficiency and naturally occurring off-axis peaking which aligns with the bootstrap current maximum. There may be other additional sources of current drive including ion cyclotron waves, helicon waves, electron cyclotron waves, and neutral beams. Typically a small amount of ion cyclotron power is utilized to fill in the current density profile and provide heat near the axis. The alternate RF sources typically have comparable but lower efficiencies than lower hybrid waves [3,35]. Also, neutral beams do not easily extrapolate into the reactor regime because of technological constraints (e.g. high density penetration problems, large size, large cost). Thus, assuming that essentially all the current is driven by lower hybrid waves is an optimistic assumption in terms of current drive efficiency.

Note that current drive power also provides heating and corresponding access to H-mode operation. However, current drive is its primary (and less efficient) mission. Therefore, although we still use the notation $P_A$ for current drive power, it should be understood that the corresponding lower hybrid waves have a carefully chosen unidirectional wavelength spectrum to maximize current drive efficiency.

The externally driven lower hybrid current $I_{CD}$(MA) is given in terms of the current drive efficiency $\eta_{CD}$(MA/MW-m$^2$), defined as follows,

$$I_{CD} = \eta_{CD} \frac{P_A}{\bar{n}_{20} R_0} \quad \text{MA} \tag{38}$$

The current drive fraction $f_{CD} = I_{CD} / I_M$ can then be written as

$$f_{CD} = \eta_{CD} \frac{P_A}{\bar{n}_{20} R_0 I_M} = \eta_{CD} \frac{P_F}{Q \bar{n}_{20} R_0 I_M} \tag{39}$$

Experiments and self-consistent, multi-dinensional ray tracing and RF simulations [3,36-39] indicate that values for $\eta_{CD}$ up to $0.3 - 0.4$ can be achieved in optimized scenarios. The underlying theory indicates that $\eta_{CD}$ is actually a function of $\bar{n}_{20}, \bar{T}_k, B_0$ and this dependence should be included in the design to obtain accurate results. However, such a self-consistent calculation of $\eta_{CD} = \eta_{CD}(\bar{n}_{20}, \bar{T}_k, B_0)$ requires considerable analysis, and is not actually necessary for compatibility with the accuracy of the rest of the analysis. Stated differently, for our purposes we shall simply assume the slightly optimistic value $\eta_{CD} = 0.35$.

- **Plasma power balance constraint**

Plasma power balance is the basic relation that determines the operating conditions for both steady state and pulsed reactors. In the context of our analysis, the relation ultimately determines the required value of $H$. The starting point for the analysis is the general time independent power balance relation given by

$$\text{Power in} = \text{Power out}$$
$$P_\alpha + P_A + P_\Omega = P_\kappa + P_R \tag{40}$$

where the undefined terms are $P_\Omega$, the Ohmic heating power and $P_R$, the radiated power, assumed to be generated primarily by Bremsstrahlung radiation. Note that except at the beginning of start-up, when the temperature is low, the Ohmic heating is small. Without much loss in accuracy, we can therefore neglect Ohmic heating for both applications of power balance. Also, during power producing operation of either type reactor, the Bremsstrahlung radiation makes a relatively small contribution to plasma power balance at typical plasma temperatures, and can be neglected [40]. This is a reasonable, although not great, approximation, but is made for analytic simplicity. Consequently, in this important regime, Eq. (40) reduces to

$$P_\alpha + P_A = P_\kappa \tag{41}$$

Expressions already have been derived for $P_\alpha$ and $P_\kappa$ in terms of the basic design variables. As is customary, it is convenient to express Eq. (41) in terms of the Lawson triple product. A short calculation using Eqs. (23)-(25) leads to

$$\bar{n}_{20}\bar{T}_k\tau_E = K_L \frac{P_\alpha}{P_\alpha + P_A} = K_L \frac{Q}{Q+5}$$

$$K_L = 17.04\left[\frac{1+f_D}{f_D^2}\left(\frac{1+2\nu_n+2\nu_T}{1+\nu_n+\nu_T}\right)\frac{1}{(1+\nu_n)(1+\nu_T)}\right]$$

(42)

For a given value of $Q$, this equation, as stated, determines $H$, which appears in $\tau_E$. The value of $H$, therefore, depends on whether we are considering a steady state or pulsed reactor, and is derived detail in Sections 6 and 7.

## 5. Nuclear and engineering constraints

In analogy with the plasma physics, there are nuclear and engineering constraints that must be satisfied for a successful reactor design. Some of these are derived quantities that can be evaluated, and shown to be satisfied, once the plasma geometry has been determined. However, other constraints can actually drive the design. The nuclear and engineering constraints are described below.

### 5.1 Nuclear constraints

- **Neutron wall loading constraint**

The neutron wall loading limit arises from the fact that all the fusion neutron power passes through the first wall surrounding the plasma. The magnitude of this power is limited by potential neutron damage to the first wall. The limit is often characterized in the literature by a maximum allowable wall loading power flux denoted by $P_W$. Typical values lie in the range $P_W \sim 2-4$ MW/m$^2$ [12,41-45].

Although convenient, this is not really the correct way to specify the limit. The reason is that the damage limit is a consequence of accumulated high energy neutron fluence rather than instantaneous power flux. Converting from fluence to flux requires a knowledge of first wall replacement time, cost of wall replacement, loss of revenue during down time, maximum acceptable output power of the plant, material properties of the wall, etc. [41]. To avoid these complications, we shall, for simplicity, require that $P_W < 2.5$ MW/m$^2$, despite its inappropriateness as a rigorously valid limit  The value of the neutron wall loading power flux is determined from

$$P_W = \frac{P_n}{S} \qquad (43)$$

where $S$ is the surface area of the first wall. The required expression for $P_W$ is obtained by making use of Eqs. (23) and (30),

$$P_W = K_W \bar{n}_{20}^2 \bar{T}_k^2 R_0 \quad \text{MW/m}^2 \qquad K_W = 2.824 \times 10^{-3} \frac{(1+\nu_n)^2(1+\nu_T)^2}{1+2\nu_n+2\nu_T} f_D^2 \varepsilon \kappa^{0.4} \qquad (44)$$

- **Minimum blanket region thickness**

The blanket region of the reactor is illustrated in Fig. 4. The three main components are: (1) the vacuum chamber, (2) the blanket, and (3) the shield. The total thickness of the blanket region is denoted by $b$. The blanket components are critical to the reactor design in terms of setting both the minimum geometric scale and corresponding capital cost.

The blanket itself has three primary functions. First, it must convert the fusion neutron energy into heat by means of slowing down collisions. Second, it must breed tritium primarily through the reaction $_3Li^6 + n(\text{slow}) \rightarrow \alpha + T + 4.8$ MeV. Third, because of unavoidable losses a thin neutron multiplier is necessary to achieve a tritium breeding ratio of $TBR = 1.1$. Just beyond the blanket is a shield whose main purpose is to limit the flux of high energy $(> 0.1 \text{ MeV})$ neutrons entering the toroidal field magnets, thereby preventing radiation damage.

The blanket region analysis is based on an examination of existing, detailed reactor designs [3,46-48] and independent MCNP simulations. There are multiple options in the analysis involving the choice of blanket materials, the geometry, the details of the cooling system, and even whether the blanket is a solid or liquid. Even so, all studies show that within a relatively small margin the most optimistic overall blanket region has a minimum thickness of approximately

$$b = 1 \text{ m} \tag{45}$$

Within this overall region, the blanket itself dominates the size of $b$, and is largely determined by the slowing down mean free path of 14.1 MeV neutrons, a basic unavoidable nuclear property. In other words there is not very much that can be done to substantially reduce $b$ from its value in Eq. (45).

It is important to keep in mind that the actual physical properties of most of the materials in an environment of fusion neutrons are at best still only marginally known from experimental data. Equally important, while any individual component may perform satisfactorily, the complex integration of an entire blanket and shield has yet to be tested experimentally, and this remains an important area of future research.

**5.2 Engineering constraints**

- **Divertor heat load constraint**

There are three ways in which power leaves the plasma core: fusion neutrons $P_n$, radiation $P_R$, and heat conduction $P_\kappa$. The first two pass through the plasma surface and are distributed more or less uniformly over the first wall surface area. The heat loss on the other hand enters the scrape-off layer, primarily at the outboard midplane, where it then flows parallel to the field. It is ultimately dissipated by a combination of localized contact with the divertor plates, approximately spatially uniform radiation resulting from detachment, and perpendicular transport of particles and energy across the scrape-off layer.

For reactor scale devices, the potential contact area with the divertor plates, even including field line spreading, is too small to dissipate the heat load by itself. When the heating power is too large, damage (sputtering, embrittlement, etc.) will occur plus the resulting divertor impurities may re-enter the plasma causing large radiation losses and degraded plasma performance. Stated differently, in a reactor environment there must be a high level of detachment to spread the heat load.

Dissipating the heat load is a very serious problem and no satisfactory solution has yet been demonstrated experimentally. In fact, today's standard divertor designs using existing materials are not satisfactory when extrapolated to a reactor – divertor heat load is a potential showstopper. There are new ideas involving extended long leg divertors [49-53], advanced materials [54], and possibly even liquid metal first walls and divertors [55], but these still need to be built and tested in high performance tokamaks.

With respect to our analysis, we acknowledge the standard assumption that a maximum heat load on the divertor plate of the order of 10 MW/m$^2$ can be satisfactorily cooled with acceptable material damage levels. However, the heat load leaving the plasma at the outer midplane is about a factor of 100 times larger. Overcoming this huge factor of 100 is the challenge for the divertor design effort.

Because divertor design is still a work in progress, our approach is to evaluate a figure of merit related to the outer midplane parallel heat flux to compare with other reactor designs. The goal is to confirm that our reactors do not lead to heat loads far in excess of other designs, keeping in mind that even if true, a solution still needs to be discovered and demonstrated experimentally. The figure of merit is derived as follows.

For a double null divertor, the poloidal component of heat flux that flows towards each of the divertor plates has the value

$$q_P = \frac{1}{2}\left(\frac{P_\kappa}{2\pi R_0 \lambda_S}\right) \qquad (46)$$

where $\lambda_S$ is the width of the scrape-off layer.

The value of $\lambda_S$ has been determined empirically from experimental measurements. A good fit to the data results in a surprisingly simple expression given by [58]

$$\lambda_S = \frac{C_S}{\overline{B}_P^{1.2}} = \frac{6.3 \times 10^{-4}}{\overline{B}_P^{1.2}} \quad \text{m} \qquad (47)$$

with $\bar{B}_P = \bar{B}_P(a)$ being the average poloidal field at the plasma edge. In general, $\lambda_S$ is quite small, on the order of millimeters.

Now, while the value of $q_P$ is critical for divertor survival, the theoretical quantity of plasma physics interest in the study of heat flow is actually the parallel heat flux $q_\parallel \approx q_\phi$. This is much larger than the poloidal heat flux by the ratio $B_\phi / \bar{B}_P$ which leads to

$$q_\parallel \approx \frac{1}{2}\left(\frac{P_\kappa}{2\pi R_0 \lambda_S}\right)\frac{B_\phi}{\bar{B}_P} \approx \frac{1}{2}\left(\frac{\bar{B}_P^{0.2}}{2\pi C_S}\right)\frac{P_\kappa B_0}{R_0} \tag{48}$$

For simplicity, the weak 0.2 power dependence of $\bar{B}_P$ can be neglected leading to an easy to evaluate heat flux figure of merit defined by

$$h_\parallel \equiv \frac{P_\kappa B_0}{R_0} = \left(\frac{Q+5}{5Q}\right)\frac{P_F B_0}{R_0} \quad \text{MW-T/m} \tag{49}$$

This is the parameter often used by the fusion community. We shall also use $h_\parallel$ to compare steady state and pulsed reactors, admitting that the "we are no worse than you are" strategy is on shaky grounds. For reference, some typical values of $h_\parallel$ for various tokamak reactor designs are as follows: ITER = 103, ARC = 400, ARIES-ACT1 = 389, ARIES-ACT2 = 568, EURO-DEMO = 287. For existing experiments without alpha power, a relatively high value is Alcator C-Mod = 25.

- **The maximum magnetic field constraint**

There is a limit to the maximum allowable magnetic field $B_{max}$ in a tokamak reactor. For the TF magnet, this field occurs on the inner leg adjacent to the blanket/shield region. In the OH transformer, $B_{max}$ is the essentially uniform field within the solenoid. For low temperature superconductors (LTS), $B_{max}$ is set by the normal-to-superconducting transition properties of the superconducting material. As an example the practical limit for a niobium-tin magnet is $B_{max} \approx 13$ T, which is the value used in the design of ITER.

However, for REBCO HTS tapes the transition limit is considerably higher, on the order of 30 T. See Fig. 5. The value is sufficiently high that it does not in general set the practical limit on $B_{max}$. Instead, it is other engineering design requirements that set the limit. For instance as the field increases, (a) substantially larger amounts of structure are needed, (b) the device becomes smaller such that the central hole space size becomes an issue, (c) joints become more complicated, etc. These real world issues lead to the conclusion that the maximum practical magnetic field in HTS magnets is approximately

$$B_{max} \approx 22 - 25 \text{ T} \qquad (50)$$

Equation (50) implies that the TF constraint on the magnetic field on the inside of the coil, denoted by $B_C$, must satisfy

$$B_C = \frac{B_0}{(1 - \varepsilon_B)} \leq B_{max} \qquad (51)$$

where $\varepsilon_B = (a + b) / R_0$. When substituting numerical values we shall set $B_{max} = 23$ T. Similarly, the essentially uniform magnetic field in the central hole of the OH transformer, $B_\Omega$, is constrained by

$$B_\Omega \leq \hat{B}_{max} \qquad (52)$$

Here also, we shall set $\hat{B}_{max} = 23$ T when numerical values are required.

- **The toroidal field magnet**

The task here is to calculate the basic geometric properties of the coils comprising the TF magnet system. Specifically, we need to determine the equivalent inboard thicknesses of each of the separate components comprising the TF magnets: the structural support material ($c_S$), superconducting tape ($c_J$), copper ($c_{CU}$), and helium cooling channels ($c_{He}$). A knowledge of these quantities allows us to determine the overall thickness $c$ of each coil, which is important in sizing the whole reactor. We again refer the reader to Fig. 1. We therefore write

$$c = c_S + c_J + c_{CU} + c_{He} \tag{53}$$

The goal is to calculate each of these quantities as a function of $B_0$ and $R_0$.

There are several steps in the analysis. First, it is necessary to calculate the amount of structural material (e.g. Inconel 718) needed to support each coil against magnetically induced stresses[2]. Second, we must calculate the required current flowing in the cable to produce the desired magnetic field on axis, thus determining the length of superconducting tape. Third, we need to calculate the amount of copper to provide protection against a partial or full quench. Lastly, we must calculate the size of the cooling channels to keep the magnet in its superconducting state.

The TF analysis requires a lengthy calculation. The details are presented in Appendix B. The end result is that the total magnet thickness can be written as

$$c = c_S + c_J + c_{CU} + c_{He} \approx c_S + 3c_J \approx c_S$$

$$c_S = \frac{R_0 B_0^2}{4\mu_0 \sigma_{max}} \left[ \frac{1}{1-\varepsilon_B} \ln\left(\frac{1+\varepsilon_B}{1-\varepsilon_B}\right) + \frac{4\varepsilon_B}{1-\varepsilon_B^2} \right] \quad (54)$$

$$c_J = \frac{B_0}{\mu_0 J_{max}} \frac{1}{1-\varepsilon_B}$$

---

[2] It is worth noting that to carry out the analysis we assume for simplicity that the centering force on each TF coil is balanced solely by wedging forces produced by adjacent coils. In practice, a bucking cylinder may be utilized for ease and reliability of the engineering. While this makes a major difference in the actual engineering design, the overall magnet thickness will not vary by very much. One way or another, a comparable amount of structural material is needed to support the magnet stresses. At the level of our analysis it is only the amount, and not the design details, that is required to size the TF magnet.

Here, the quantity $\sigma_{max} \approx 650$ MPa is the maximum allowable mechanical stress that can be supported by the structural material while $J_{max} \approx 700$ A/mm$^2$ is the maximum current density that can flow in the HTS superconducting tapes. For high field magnets the typical situation has $3c_J \ll c_S$ allowing us to set $c \approx c_S$ corresponding to the final form in Eq. (54).

- **The OH transformer**

The OH transformer is a vertically oriented, cylindrically symmetric solenoid as illustrated in Fig. 1. It is often referred to as the central solenoid (CS). Usually the OH transformer is segmented, but for our purposes, it is sufficient to treat it as a single long solenoid. As with the TF coils, our goal is to calculate the dimensions of the transformer to help size the overall reactor. This is a more critical calculation for pulsed systems since the central hole in the transformer must be large enough to provide a sufficient volt-second swing to sustain the plasma for greater-than-an-hour flat top pulses. For a steady state reactor with current drive the transformer demands are much smaller – sufficient volt-seconds are required only to raise the plasma current from zero to its final desired operating value. This is a small fraction of the total requirement in a pulsed reactor.

The pulse length that must be provided by the OH transformer is an important driving constraint in the design of pulsed reactors. Its value is determined by three main factors: (1) the number of allowable cycles before replacement is needed, (2) the OH replacement down time, and (3) the need for high average power (i.e. high duty factor). The pulse length basically determines the radius $R_\Omega$ of the transformer. The thickness of the solenoid $d$ is determined by a combination of structural support requirements and current carrying capacity. The analysis is similarly complicated as that of the TF coils and is presented in Appendix C. The main results can be summarized as follows.

**Height**

$$L_\Omega = 2(\kappa a + b + c) \tag{55}$$

**Radius + thickness**

$$R_\Omega + d = \left\{ \frac{K_\Omega D^2(\zeta)}{(2\mu_0 \hat{\sigma}_{\max})^{1/2}} \left[ 1 + \frac{R_0^2}{(\kappa a + b + c)^2} \right]^{1/2} \left( 1 - f_B + \frac{\tau_{L/R}}{\tau_P} \right) \frac{I_M}{R_0 \bar{T}_k^{3/2}} \right\}^{1/2}$$

$$K_\Omega = 34.38 \frac{\tau_P}{\varepsilon^2 \kappa G}$$

$$G = (1 + \nu_T)^{3/2} \left[ \frac{1}{\nu_0} - 2\varepsilon^{1/2} \frac{\Gamma(\nu_0)\Gamma(5/4)}{\Gamma(\nu_0 + 5/4)} + \varepsilon \frac{\Gamma(\nu_0)\Gamma(3/2)}{\Gamma(\nu_0 + 3/2)} \right] \quad \nu_0 = 1 + (3/2)\nu_T$$

$$D = \left[ \frac{1}{\zeta^{1/2}} \frac{(1 + \zeta)^2}{1 + \zeta + \zeta^2/3} \right]^{1/2} \quad \zeta = \frac{B_\Omega^2}{2\mu_0 \hat{\sigma}_{\max}} = \frac{\hat{B}_{\max}^2}{2\mu_0 \hat{\sigma}_{\max}}$$

$$f_B = K_b \frac{\bar{n}_{20} \bar{T}_k R_0^2}{I_M^2}$$

$$\frac{\tau_{L/R}}{\tau_P} = K_\tau \left[ \ln\left(\frac{8}{\varepsilon}\right) - 2 \right] R_0^2 \bar{T}_k^{3/2} \quad K_\tau = 5.818 \times 10^{-3} \frac{\varepsilon^2 \kappa G}{\tau_P} \tag{56}$$

Here, $B_\Omega \propto \zeta^{1/2}$ is the approximately uniform field in the center of the OH transformer. Its only appearance in the overall analysis is in the function $D(\zeta)$ which, as is shown in Appendix C, is a rapidly decreasing function of $\zeta$. Therefore, to minimize the OH size $R_\Omega + d$ (and corresponding reactor cost) we must choose $B_\Omega$ as large as possible. While this may be intuitively clear, it is directly proven by the analysis, thereby confirming the hypothesis that high field can have a positive impact on compact reactor design. We thus set $B_\Omega = \hat{B}_{\max} \approx 23$ T to minimize cost.

Also, $\hat{\sigma}_{max} \approx 500$ MPa is the maximum allowable values as set by technology. In evaluating the OH quantities we have assumed that the transformer is constructed of the same HTS and structural material as the TF coils. However, we require $\hat{\sigma}_{max} < \sigma_{max}$ to improve CS survival due to the cyclical nature of the stresses. The pulse length $\tau_P$ is measured in hours. Its value is approximately $\tau_P \approx 1.5$ hours as determined by the relation

$$\tau_P = 720 \frac{N_{POW} N_{REP}}{N_{CYC}} \approx 1.5 \quad \text{hours per pulse} \tag{57}$$

where $N_{CYC} \approx 30,000$ is the number of operating cycles before OH replacement is needed, $N_{REP} \approx 6$ months is the number of months required to replace the OH transformer, and $N_{POW} \approx 10$ is the number of replacement periods during which the reactor is operating and producing power. The value 10 gives a reasonably high average power over a full operating-replacement cycle.

- **The Ohmic current constraint**

As stated, the Ohmic current constraint is a dominant driver in pulsed reactors. The constraint requires that the major radius be sufficiently large so that the TF magnet, blanket, and OH transformer all be able to fit within the central hole of the reactor. Since the OH transformer radius $R_\Omega$ is relatively large in order to provide the required flux swing, this leads to an important constraint on the minimum size of $R_0$. The constraint is determined by the need for the Ohmic current fraction $f_\Omega$ to satisfy the current generation requirement, $f_\Omega = 1 - f_B$. It can be written as

$$R_0 = a + b + c + d + R_\Omega \tag{58}$$

- **The cost metric**

A key parameter in the steady state versus pulsed reactor comparison is the magnetic energy cost metric, repeated here for convenience.

$$C_{MAG} = \frac{W_{TF}}{P_F} \quad \text{MJ/MW} \tag{59}$$

The toroidal magnetic energy within the TF magnet volume is approximately given by

$$W_{TF} \approx \frac{B_0^2}{2\mu_0} V_{TF} = \frac{B_0^2}{2\mu_0}(2\pi R_0)[2(\kappa a + b)][2(a+b)] = K_{TF} R_0^3 B_0^2 \quad \text{MJ}$$

$$K_{TF} = 10\left(1 + \frac{b}{\kappa \varepsilon R_0}\right)\left(1 + \frac{b}{\varepsilon R_0}\right)\varepsilon^2 \kappa \tag{60}$$

Therefore,

$$C_{MAG} = K_C R_0^3 B_0^2 \quad \text{MJ/MW} \qquad K_C = \frac{K_{TF}}{P_F} \tag{61}$$

The quantity $C_{MAG}$ is a measure of the capital cost per watt of fusion energy.

Recall that there is a secondary inverse cost metric $P_{VOL}$ corresponding to the plasma power density. It is equal to the ratio of the fusion power to the plasma volume, and has the value

$$P_{VOL} = \frac{P_F}{V_P} = \frac{P_F}{(2\pi R_0)(\pi a^2 \kappa)} = \frac{K_V}{R_0^3} \quad \text{MW/m}^3 \qquad K_V = 0.0507 \frac{P_F}{\varepsilon^2 \kappa} \tag{62}$$

## 6. Reactor design analysis

All of the important physics, nuclear, and engineering constraints, have now been defined. As such, we are in a position to select the most stringent subset of constraints and use them to design steady state and pulsed tokamak reactors. The end goal of the analysis is to derive values for the six basic design variables, $B_0$, $R_0$, $I_M$, $\bar{n}_{20}$, $\bar{T}_k$, $H$ for each reactor type.

The analysis proceeds in four steps. First, the most stringent set of constraints for each reactor type is defined. Second, the mathematical design analysis is described, leading to analytic algebraic expressions for the basic design variables. Third, the analysis is applied to the ARC and European Demo experiments to test the reliability of the model. Fourth, the analysis is used to design a steady state and pulsed power reactor, each with a high fusion gain. Several critical inputs to the final design are then varied to test sensitivity.

## 6.1 Driving constraints

The most stringent driving constraints are based on an examination of many more sophisticated designs in the literature [3,12,60] as well as physical intuition. The choices made are self-correcting in that if a wrong choice is made, one of the unused constraints will be violated a posteriori and the analysis will have to be redone.

Since there are six basic design variables, we must choose the six most stringent constraints to carry out the design. Five of the constraints overlap for each reactor and are listed in Table 6.1.

| Constraint | Steady State and Pulsed |
|---|---|
| **TF field relation** | $\dfrac{B_0}{1-\varepsilon_B} = B_C$ |
| **Greenwald density limit** | $\dfrac{\bar{n}_{20} R_0^2}{I_M} = K_G$ |
| **Fusion power mission** | $\bar{n}_{20}^2 \bar{T}_k^2 R_0^3 = \dfrac{P_F}{K_F}$ |
| **MHD Troyon beta limit** | $\dfrac{\bar{n}_{20} \bar{T}_k R_0}{I_M B_0} = K_\beta$ |
| **Plasma power balance** | $\dfrac{P_\alpha + P_A}{P_\alpha} \bar{n}_{20} \bar{T}_k \tau_E = K_L$ |

Table 6.1 The five overlapping constraints for a steady state and pulsed reactor.

Observe that both steady state and pulsed reactors must satisfy the same TF field relation (with $B_C \leq B_{\max}$), Greenwald density limit, fusion power mission, Troyon $\beta$ limit, and plasma power balance constraints. The one difference in the constraints is the method of current generation for each reactor. For a steady state reactor sufficient current drive is required to generate the required current. For a pulsed reactor the OH transformer radius must be large enough to generate a sufficiently long pulse at the required current. This difference is shown in Table 6.2.

| Constraint | Steady State | Pulsed |
|---|---|---|
| **Current generation (CD vs. OH)** | $f_B + f_{CD} = 1$ | $R_0 = a + b + c + d + R_\Omega$ |

Table 6.2 The different current generation constraints for each reactor type

## 6.2 Mathematical design analysis

The mathematical design analysis is described as follows. First, after straightforward algebra the TF field relation, Greenwald density limit, fusion power mission, and MHD Troyon $\beta$ limit, constraints lead to expressions for $\bar{n}_{20}, \bar{T}_k, I_M, B_0$ as functions of $R_0$ and $B_C$. Second, after slightly tedious algebra, plasma power balance yields an expression for $H$, also as a function of $R_0$ and $B_C$. Lastly, application of the current generation constraint yields a relationship between $R_0$ and $B_C$ for each reactor type. The end result is a set of analytic algebraic expressions for each of the basic design variables.

- **Expressions for $\bar{n}_{20}, \bar{T}_k, I_M, B_0$**

Simple algebraic elimination leads to expressions for $\bar{n}_{20}, \bar{T}_k, I_M, B_0$ as functions of $R_0$ and $B_C$. These expressions are obtained from the five constraints listed in Table 6.1. The results are the same relations for both steady state and pulsed reactors.

$$B_0 = B_C\left(1 - \varepsilon - \frac{b}{R_0}\right)$$

$$I_M = \frac{K_I}{B_0 R_0^{1/2}} \qquad K_I = \frac{1}{K_\beta}\left(\frac{P_F}{K_F}\right)^{1/2}$$

$$\bar{n}_{20} = \frac{K_n}{B_0 R_0^{5/2}} \qquad K_n = \frac{K_G}{K_\beta}\left(\frac{P_F}{K_F}\right)^{1/2} \qquad (63)$$

$$\bar{T}_k = K_T B_0 R_0 \qquad K_T = \frac{K_\beta}{K_G}$$

For notational compactness, we have expressed the quantities of interest in terms of $B_0$ which is simply related to $R_0$ and $B_C$ through the top equation in Eq. (63).

- **Evaluation of $H$**

The fusion gain $H = H(R_0, B_C)$ can be evaluated by using the plasma power balance relation with $\tau_E$ given by Eq. (26). A short calculation shows that this relation can be written as

$$\left(\frac{Q}{Q+5}\right)^{0.31} = \left[0.4402 \frac{A^{0.19} \varepsilon^{0.58} \kappa^{0.78}}{K_L P_F^{0.69}}\right] H R_0^{1.97} B_0^{0.15} I_M^{0.93} \bar{n}_{20}^{1.41} \bar{T}_k \qquad (64)$$

and again is valid for both steady state and pulsed reactors. The terms involving $I_M, \bar{n}_{20}, \bar{T}_k$ have already been expressed as functions of $R_0$ and $B_0$ for both reactor types in Eq. (63). It is then a straightforward, although slightly tedious calculation, to substitute these functions into Eq. (64) and solve for $H$. The result is

$$H = K_H B_0^{1.19} R_0^{1.02} \qquad K_H = 2.272 \frac{K_L K_\beta^{1.34} K_F^{1.17}}{A^{.19} \varepsilon^{0.58} \kappa^{0.78} K_G^{0.41} P_F^{0.48}} \left(\frac{Q}{Q+5}\right)^{0.31} \qquad (65)$$

Observe that large changes in $Q$, particularly for high $Q$, tend to produce only small changes in $H$. This is the source of the problem alluded to earlier that motivates the change from $Q$ to $H$ as one of the basic design variables. The sensitivity can ultimately be traced back to the $(P_\alpha + P_A)^{0.69}$ dependence in $\tau_E$.

- **Calculating $R_0$**

As the analysis now stands, the quantities $B_0, I_M, \bar{n}_{20}, \bar{T}_k, H$ have all been expressed in terms of $R_0$ and $B_C$, the relations being the same for both reactor types. The system is closed by applying the appropriate current generation constraint, which yields an algebraic equation of the form $F(R_0, B_C) = 0$. This equation is different for each reactor type. It is useful to think of inverting $F(R_0, B_C) = 0$ leading to $R_0 = R_0(B_C)$. Varying $B_C$ should then demonstrate whether the basic intuition of setting $B_C = B_{\max}$ leads to the smallest $R_0$ and cost.

Determining the functions $F(R_0, B_C)$ for each reactor type is straightforward since all the required terms have been already evaluated. As above, the results can be expressed in compact form, $F(R_0, B_C) = 0 \rightarrow F(R_0, B_0) = 0$, by recalling that

$$B_C(B_0, R_0) = \frac{B_0}{1 - \varepsilon_B} \qquad \varepsilon_B = \varepsilon + \frac{b}{R_0} \tag{66}$$

The current drive constraint $F(R_0, B_0) = 1 - f_B - f_{CD} = 0$ for a steady state reactor is determined by substituting from Eqs. (34) and (39). The result can be written as

**Steady state current drive constraint**

$$B_0 = \left( \frac{1}{K_B R_0^{3/2} + K_{CD} R_0^2} \right)^{1/2}$$

$$f_B = K_b \frac{\bar{n}_{20} \bar{T}_k R_0^2}{I_M^2} = K_B B_0^2 R_0^{3/2} \qquad K_B = \frac{K_b K_n K_T}{K_I^2} \tag{67}$$

$$f_{CD} = \frac{\eta_{CD} P_A}{\bar{n}_{20} I_M R_0} = K_{CD} B_0^2 R_0^2 \qquad K_{CD} = \frac{\eta_{CD} P_F}{K_n K_I Q}$$

A similar, although slightly more complicated calculation, can be used to evaluate the Ohmic transformer constraint $F(R_0, B_0) = R_0 - a - b - c - d - R_\Omega = 0$. In this case, the relevant equations are Eqs. (34), (54), and (56). The Ohmic transformer constraint reduces to

**Pulsed Ohmic transformer constraint**

$$1 - \varepsilon_B - \frac{B_0^2}{4\mu_0 \sigma_{\max}} \left[ \frac{1}{1-\varepsilon_B} \ln\left(\frac{1+\varepsilon_B}{1-\varepsilon_B}\right) + \frac{4\varepsilon_B}{1-\varepsilon_B^2} \right]$$

$$= K_{OH} \left[ 1 + \frac{R_0^2}{(\kappa a + b + c)^2} \right]^{1/4} \left( 1 - f_B + \frac{\tau_{L/R}}{\tau_P} \right)^{1/2} \frac{1}{B_0^{5/4} R_0^{5/2}}$$

$$K_{OH} = \left[ \frac{K_\Omega K_I D^2}{K_T^{3/2} (2\mu_0 \hat{\sigma}_{\max})^{1/2}} \right]^{1/2} \qquad (68)$$

$$f_B = K_B B_0^2 R_0^{3/2}$$

$$\frac{\tau_{L/R}}{\tau_P} = K_\tau K_T^{3/2} \left[ \ln\left(\frac{8}{\varepsilon}\right) - 2 \right] B_0^{3/2} R_0^{7/2}$$

This completes the mathematical analysis for both types of reactor. The final result is a set of analytic algebraic expressions for each of the basic design variables as determined by the strictest set of driving constraints.

## 7. Results and Comparisons

In this Section we use our models for steady state and pulsed tokamaks to obtain results in three main categories. First, we apply our models to more sophisticated existing designs to test the accuracy of our predictions. Specifically, we make a comparison with ARC for a steady state reactor and the European Demo for a pulsed reactor. Second, the models are used to determine reference designs for each type of reactor. This enables a meaningful comparison between steady state and pulsed reactors. Third, we vary a number of key physical parameters describing each reference design to test sensitivities. This sheds light on which areas of plasma physics and engineering have high leverage in improving the attractiveness of fusion reactors.

## 7.1 Comparisons with existing designs

- **ARC – a steady state pilot plant**

The ARC reactor [3] is a 500 MW (thermal) steady state pilot plant that makes use of the recently developed REBCO superconducting tapes, as well as several other engineering innovations. Its mission is similar to the stated goals of our power reactor except that as a pilot plant its $Q$ value is about one-half the minimum requirement of a commercial plant. In applying our model, we provide as inputs from the ARC design the same set of parameters that we will use when designing the reference reactor. A comparison of the predictions then provides a basis for assessing the reliability of our simplified model.

The inputs for the ARC comparison are listed below in Table 7.1.

| Parameter | Symbol | Value |
|---|---|---|
| Greenwald density limit | $N_G$ | 0.67 |
| Elongation | $\kappa$ | 1.84 |
| Inverse aspect ratio | $\varepsilon$ | 0.342 |
| Beta limit | $\beta_N$ | 0.0259 |
| Fusion gain | $Q$ | 13.6 |
| Fuel dilution factor | $f_D$ | 0.85 |
| Current drive efficiency | $\eta_{CD}$ | 0.321 |
| Thermal fusion power (MW) | $P_F$ | 525 |
| Maximum field on the TF (T) | $B_C$ | 23 |
| Blanket/shield thickness (m) | $b$ | 0.85 m |
| Number density profile factor | $\nu_n$ | 0.385 |
| Temperature profile factor | $\nu_T$ | 0.929 |
| Current density profile factor (approximate) | $\nu_J$ | 0.472 |
| Thermal conversion efficiency | $\eta_T$ | 0.4 |

Table 7.1 Input parameters from the published ARC design

These parameters are used as inputs to our steady state model. It is then a simple matter to predict the basic design variables as well as several other parameters of interest. Our predictions are listed in Table 7.2 along with the actual ARC values, in order to make comparisons.

| Parameter | Symbol | ARC Value | Model Value |
|---|---|---|---|
| Major radius (m) | $R_0$ | 3.3 | 3.38 |
| Plasma magnetic field (T) | $B_0$ | 9.2 | 9.35 |
| Plasma current (MA) | $I_M$ | 7.8 | 7.93 |
| Average density ($10^{20}\,\mathrm{m}^{-3}$) | $\bar{n}_{20}$ | 1.3 | 1.27 |
| Average temperature (keV) | $\bar{T}_k$ | 14 | 14.1 |
| Confinement enhancement factor | **H** | **1.8** | **1.86** |
| Bootstrap fraction | $f_B$ | 0.63 | 0.635 |
| Electric power out (MWe) | $P_E$ | 283 | 267 |
| Recirculating power fraction | $f_{RP}$ | 0.273 | 0.289 |
| Absorbed RF power (MW) | $P_A$ | 38.6 | 38.6 |
| Kink safety factor | $q_*$ | 4.99 | 5.05 |
| Heating power/LH threshold | $(P_\alpha + P_A)/P_{LH}$ | 3.42 | 3.31 |
| Neutron wall loading (MW/m$^2$) | $P_W$ | 1.98 | 1.89 |
| Heat flux parameter (MW-T/m) | $h_\parallel$ | 400 | 397 |
| TF magnetic energy (GJ) | $W_{TF}$ | 18 | 17.6 |
| Magnetic energy metric (MJ/MW) | $C_{MAG}$ | 34.3 | 35.2 |
| Power density metric (MW/m$^3$) | $P_{VOL}$ | 3.72 | 3.21 |

Table 7.2

Comparison of ARC parameters with those predicted by our model

The comparison shows that there is surprisingly good agreement between our simple model and the actual ARC design. The overall good agreement is not so much a consequence of "lucky coincidence" or "brilliant mathematics" but, as stated previously, the result of choosing the proper set of most stringent constraints. Observe that all the remaining unused constraints lie within the acceptable range, except for $H$ which is nearly double its allowable value.

- **European Demo – A pulsed demonstration power plant**

The European Demo is a pulsed demonstration power plant that utilizes existing technology and relatively conservative plasma physics [13,45,61,62]. It is a large plant, producing about 1000 MWe. It makes use of existing LTS superconducting magnet technology, which limits the maximum field to about 13 T. The decision to use pulsed rather than steady state technology is presumably based on the judgement that RF current drive has low efficiency and high cost, plus is not as reliable technologically as Ohmically driven current.

As for the ARC analysis, we shall specify the same set of Demo inputs that will be used in the pulsed reference design. Our model then predicts values for the basic design variables plus other quantities of interest, which can be compared with the actual Demo design. We acknowledge that some of the input parameters are a little more uncertain than with ARC. Also, Demo does use a small amount, (less than 10%), of neutral beam (NB) current drive to optimize the design, which we neglect in our comparison. On the one hand, our model shows that including it makes only a small difference in the predictions. On the other hand, by ignoring the NB current drive, we are using the exact same model for the comparison and for our reference design, and this is the path we have chosen.

The input parameters for the European Demo are listed in Table 7.3.

| Parameter | Symbol | Value |
|---|---|---|
| Greenwald density limit | **$N_G$** | **1.2** |
| Elongation | $\kappa$ | 1.59 |
| Inverse aspect ratio | $\varepsilon$ | 0.323 |
| Beta limit | $\beta_N$ | 0.0259 |
| Fusion gain | $Q$ | 39.9 |
| Fuel dilution factor | $f_D$ | 0.775 |
| Thermal fusion power (MW) | $P_F$ | 2037 |
| Maximum field on the TF (T) | $B_{max}$ | 12.3 |
| Maximum field on the CS (T) | $\hat{B}_{max}$ | 12.9 |
| Maximum stress on the TF (MPa) | $\sigma_{max}$ | 660 |
| Maximum stress on the CS (MPa) | $\hat{\sigma}_{max}$ | 660 |
| Flat top pulse length (hours) | $\tau_P$ | 2 |
| Blanket/shield thickness (m) | $b$ | 1.63 m |
| Number density profile factor | $\nu_n$ | 0.27 |
| Temperature profile factor | $\nu_T$ | 1.094 |
| Thermal conversion efficiency | $\eta_T$ | 0.4 |

Table 7.3

Input parameters for the European Demo

Observe that all of the input parameters are within the allowable range except for the Greenwald density fraction $N_G$ which is slightly above the maximum limit $N_G = 1$.

Straightforward application of the pulsed tokamak model leads to the parameter predictions listed in Table 7.4. Also listed are the actual Demo values.

| Parameter | Symbol | Demo Value | Model Value |
|---|---|---|---|
| Major radius (m) | $R_0$ | 9.07 | 8.09 |
| Plasma magnetic field (T) | $B_0$ | 5.67 | 5.85 |
| Plasma current (MA) | $I_M$ | 19.6 | 18.9 |
| Average density ($10^{20}\,\mathrm{m}^{-3}$) | $\overline{n}_{20}$ | 0.798 | 1.06 |
| Average temperature (keV) | $\overline{T}_k$ | 13.1 | 11.4 |
| **Confinement enhancement factor** | **$H$** | **1.1** | **0.93** |
| Bootstrap fraction | $f_B$ | 0.348 | 0.179 |
| Current density profile factor | $\nu_J$ | 0.176 | 0.233 |
| Total electric power out (MWe) | $P_E$ | 914 | 1037 |
| Absorbed NB power (MW) | $P_A$ | 50 | 51.1 |
| Kink safety factor | $q_*$ | 2.71 | 2.35 |
| Heating power/LH threshold | $(P_\alpha + P_A)/P_{LH}$ | 3.93 | 3.88 |
| Neutron wall loading (MW/m$^2$) | $P_W$ | 1.05 | 1.48 |
| Heat flux parameter (MW-T/m) | $h_\parallel$ | 287 | 331 |
| TF magnetic energy (GJ) | $W_{TF}$ | 136 | 67.9 |
| Magnetic energy metric (MJ/MW) | $C_{MAG}$ | 66.6 | 33.3 |
| Power density metric (MW/m$^3$) | $P_{VOL}$ | 0.814 | 1.18 |

Table 7.4

Comparison of Demo parameters with those predicted by our model

Again, there is reasonably good agreement between the actual Demo design and our pulsed reactor model. Note that the actual Demo design parameters all lie within an allowable range except for the confinement enhancement factor which is slightly above unity: $H = 1.1$ Our model predicts $H = 0.91$. The values are close, with our model being slightly more optimistic, placing the design on the safe side of the curve. The Demo value is much closer to empirical value $H = 1$ than the steady state design which requires $H = 1.8$.

There is a substantial difference in the prediction of the bootstrap fraction, with our model being more pessimistic. The main reason is that we use our model density profile to calculate $f_B$ which, from the Demo data, has a small value of $\nu_n$. The corresponding density is quite flat leading to a low value of $f_B$. The actual Demo design uses a more realistic density profile, with a steep edge pedestal, when calculating $f_B$, leading to a higher value. Even so, unlike the steady state model, the bootstrap current is only a relatively small fraction of the total current, so that it has a small effect on the overall design.

The other main difference is in the stored TF magnetic energy, even though the central fields $B_0$ are similar. This difference is likely due to the fact that Demo has additional TF volume on the outboard side of the plasma in order to keep the ripple to an acceptable level. This extra volume requires additional stored magnetic energy.

The main conclusion from the comparisons is as follows. Although ARC and Demo are widely different devices in terms of power, magnetic field, and method of current generation, the predictions of our simple model are in reasonably good agreement with the more accurate, sophisticated designs. This provides confidence that we can design two comparable reactors and make a fair comparison, which is the task of the next section.

**7.2 Steady state and pulsed reference designs**

- **Steady state reference design**

We now define a set of reference input parameters that are used to design a steady state power reactor. Most of these parameters are identical to the ones that are used for the pulsed reactor. The only differences involve parameters directly related to the method of current generation. The goal is to make the comparison as fair as possible.

To begin, note that the steady state reactor and ARC have many similarities. The main difference is the need for a higher fusion gain: $Q$ must increase from 13.6 to about 26. ARC is aimed at producing a reasonable amount of net power with a minimum capital cost, and not being overly concerned about a low recirculating power. The steady state reactor's mission is to produce the same amount of power but with a lower recirculating power to improve economics over the long operational life of the plant.

There are two steps in the analysis. First, the value of $B_C$ is allowed to float. We can then plot the dependence of various quantities of interest, such as $C_{MAG}$, $P_{VOL}$, $R_0$ and $H$ versus $B_C$ to test the hypothesis that the highest possible magnetic field is the "best" option for steady state tokamak reactors. This indeed turns out to be the case. Second, the value of $B_C = B_{max}$ is set to 23 T. The model is then used to determine the actual reference design.

A short discussion is warranted concerning the definition of "best" reactor. An obvious choice is to minimize the magnetic energy cost metric $C_{MAG}$. However, as previously stated, all the designs considered require confinement enhancement factors that exceed the limit $H = 1$. This implies the need for improved plasma physics performance, a task that may be quite difficult, based on past history. However, achieving higher field also requires improved magnet development using the new HTS superconducting tapes. It is the authors' belief that developing high field magnets has at least as high a probability of success as substantially improving energy confinement. Consequently, when conflicting choices for "best" reactor design arise, we shall assume that minimizing $H$ takes precedence over minimizing $C_{MAG}$.

With this introduction, we now define the input parameters for the steady state reference reactor, which are given in Table 7.5

| Parameter | Symbol | Value |
|---|---|---|
| Greenwald density limit | $N_G$ | 0.85 |
| Elongation | $\kappa$ | 1.8 |
| Inverse aspect ratio | $\varepsilon$ | 0.25 |
| Beta limit | $\beta_N$ | 0.026 |
| Fusion gain | $Q$ | 26 |
| Fuel dilution factor | $f_D$ | 0.85 |
| Current drive efficiency | $\eta_{CD}$ | 0.35 |
| Maximum allowable TF field (T) | $B_{\max}$ | 23 |
| Thermal fusion power (MW) | $P_F$ | 500 |
| Blanket/shield thickness (m) | $b$ | 1 m |
| Number density profile factor | $\nu_n$ | 0.4 |
| Temperature profile factor | $\nu_T$ | 1.1 |
| Current density profile factor | $\nu_J$ | 0.453 |
| Thermal conversion efficiency | $\eta_T$ | 0.4 |
| Wall to RF conversion efficiency for LHCD | $\eta_{RF}$ | 0.5 |

Table 7.5

Input parameters for the steady state reference design

Several comments are in order. Observe that the Greenwald fraction is slightly more optimistic while the blanket thickness is slightly more conservative. Also, for most of the parameters it is intuitively clear whether they should be set to their maximum or minimum allowable values. One exception is the inverse aspect ratio $\varepsilon$. The results, which are not immediately obvious, show that large aspect ratio (i.e. small $\varepsilon$) reduces the required $H$. However, it increases the cost $C_{MAG}$. This is one situation where minimizing $H$ takes precedence over minimizing $C_{MAG}$. Further discussion is presented in the sensitivity studies subsection.

The input parameters in Table 7.5 are substituted into our model. As stated we first allow the value of $B_C$ to float to see if high field is indeed the path to the "best" reactor. The relevant results are illustrated in Fig. 6 where we have plotted $R_0, H, C_{MAG}, P_{VOL}, q_*$ versus $B_C$.

Observe that the radius $R_0$ and cost $C_{MAG}$ decrease relatively rapidly with increasing $B_C$. Similarly the power density metric $P_{VOL}$ increases with increasing $B_C$ while the confinement factor decreases, but more slowly. As anticipated, its magnitude is considerably above the empirical limit $H = 1$. The kink safety factor increases slightly with $B_C$ and is always substantially above the conservative value for the stability limit, $q_* = 3$. For the steady state reactor each of these important scaling relations indicate that high field leads to the best reactor. The improved performance expected and predicted with high field confirms the conjecture that the development of REBCO HTS tapes may be a game changer for steady state tokamak reactors. We now proceed by setting $B_C = B_{\max} = 23 \text{ T}$, the maximum allowable value.

Using this value, our model predicts the following parameters for the steady state reference design, listed in Table 7.6.

| Parameter | Symbol | Model Value |
|---|---|---|
| Maximum field on the TF (T) | $B_C$ | 23 |
| Major radius (m) | $R_0$ | 4.10 |
| Minor radius (m) | $a$ | 1.02 |
| Plasma magnetic field (T) | $B_0$ | 11.6 |
| Plasma current (MA) | $I_M$ | 5.53 |
| Average density ($10^{20}\,\mathrm{m}^{-3}$) | $\overline{n}_{20}$ | 1.43 |
| Average temperature (keV) | $\overline{\overline{T}}_k$ | 12.1 |
| Confinement time (sec) | $\tau_E$ | 1.15 |
| **Confinement enhancement factor** | **$H$** | **1.94** |
| Kink safety factor | $q_*$ | 5.68 |
| On axis safety factor $q(\rho=0)$ | $q_0$ | 4.75 |
| Edge safety factor $q(\rho=1)$ | $q_a$ | 5.68 |
| Minimum safety factor $q(\rho=\rho_{\min})$ | $q_{\min}$ | 4.09 |
| Minimum $q$ normalized radius | $\rho_{\min}$ | 0.629 |
| Bootstrap fraction | $f_B$ | 0.792 |
| Current drive fraction | $f_{CD}$ | 0.208 |
| Heating power/LH threshold | $(P_\alpha + P_A)/P_{LH}$ | 2.00 |
| Electric power out (MWe) | $P_E$ | 255 |
| LHCD wall power (MWe) | $P_{RF}$ | 38.5 |
| LHCD power absorbed (MW) | $P_A$ | 19.2 |
| Recirculating power fraction | $f_{RP}$ | 0.151 |
| Neutron wall loading (MW/m$^2$) | $P_W$ | 1.70 |
| Heat flux parameter (MW-T/m) | $h_\parallel$ | 339 |
| Stored TF magnetic energy (GJ) | $W_{TF}$ | 31.9 |
| Magnetic energy metric (MJ/MW) | $C_{MAG}$ | 63.8 |
| Power density metric (MW/m$^3$) | $P_{VOL}$ | 3.29 |

Table 7.6

Output parameters for the steady state reference reactor

Observe that the parameters are similar to those of ARC. Also, all of the relevant constraints have been satisfied except for the confinement enhancement factor which has the value $H = 1.94$. This value is slightly higher than the ARC value because of the need for increased gain, from $Q = 13.6$ to $Q = 26$. The need for higher $H$ remains an important problem requiring substantially improved plasma performance. The reactor has a slightly larger major radius than ARC because of the smaller $\varepsilon$. The central magnetic field in the plasma is also larger, again because of the smaller $\varepsilon$, corresponding to a weaker $1/R$ decay of the field. The current is smaller leading to a perhaps uncomfortably larger bootstrap fraction. High bootstrap fraction is good to the extent that we can "trust" the plasma to behave reliably. The safety factor is high and the neutron wall loading is less than the 2.5 MW/m$^2$ constraint. The midplane heat flux parameter is large as expected and comparable to other reactor designs. As stated, this is still an unsolved problem.

The magnetic energy cost parameter $C_{MAG}$ is a about a factor of 2 larger than for ARC because of the need for higher gain, leading to a larger major radius and corresponding increased magnetic energy. Overall, the production of 255 MWe from a 4.1 meter tokamak reactor may be an acceptable design. The biggest plasma physics problem is the need for an enhanced value of $H$. The cost metric will be compared with that of the pulsed reactor shortly.

- **Steady state sensitivity studies**

The parameters describing the steady state reference reactor have now been defined. In this subsection, we shall investigate the sensitivity of the design to several of the input parameters. Because of the simplicity of the model, it is an easy matter to generate enormous amounts of scaling information. To limit this information, we shall restrict our studies to those parameters whose reference values are either non-intuitive, are changed because of a redefined mission, or are modified by an improvement in plasma physics performance directly related to steady state operation.

The specific parameters of interest are (a) the inverse aspect ratio $\varepsilon$, (b) the fusion power out $P_F$, and (c) the current drive efficiency $\eta_{CD}$. The analysis is straightforward. The magnetic field $B_C$ is set to its maximum value of 23 T. All other parameters are held fixed except for the sensitivity parameter under consideration. This parameter is then scanned over a reasonable range and the results plotted in a similar style to Fig. 6. Conclusions are then drawn.

We start with the inverse aspect ratio $\varepsilon$. Illustrated in Fig. 7 are curves of $R_0, H, C_{MAG}, P_{VOL}, q_*$ versus $\varepsilon$. Observe that there is a weak minimum in $R_0$ because of the competition between a higher Troyon $\beta$ limit and a weaker central field $B_0$ as $\varepsilon$ increases from the reference value. The minimum is weak because $\varepsilon$ is only allowed to vary over a relatively narrow range about the minimum. The safety factor remains high. The cost weakly, and the power density strongly, decreases with $\varepsilon$ while the required $H$ increases. In accordance with our definition of "best" reactor, we see that setting $\varepsilon$ to its minimum allowable value $\varepsilon = 0.25$ is the most desirable choice. This value is also used in the ARIES-ACT studies [12].

Why does increasing $\varepsilon$ require a larger $H$? The reason is as follows. For simplicity assume $B_0$ and $R_0$ are approximately constant (since $B_C = B_{\max}$ is fixed and $R_0$ has a weak minimum). Then, power balance shows that $H \propto 1/(I_M^{0.93} \overline{n}_{20}^{0.41} \overline{p})$. Now, the fusion power constraint shows that $\overline{p} \propto 1/\varepsilon$, which, from the Troyon beta limit, implies that $I_M \propto \varepsilon^0$ is independent of $\varepsilon$. Lastly, the Greenwald density limit shows that $\overline{n}_{20} \propto 1/\varepsilon^2$. Combining these scaling relations leads to $H \propto \varepsilon^{1.26}$ - small $\varepsilon$ leads to the smallest required confinement enhancement. Qualitatively, it is the strong inverse $a^2$ dependence of the Greenwald density limit that dominates the scaling. Large aspect ratio tokamaks have a higher density limit at fixed $B_0, R_0$.

Consider next the scaling with $P_F$. We have chosen $P_F = 500$ MW as our reference case based on the current belief that this is about as high a value as would be desirable by US industry. However, there is no plasma physics reason why $P_F$ cannot be higher, so let us assume that this becomes more acceptable in the future. Curves of $R_0, H, C_{MAG}, P_{VOL}, q_*$ versus $P_F$ are plotted in Fig. 8. The curves show that, as expected, $R_0$ increases as the output $P_F$ increases. Also expected, $C_{MAG}$ decreases and $P_{VOL}$ increases with increasing $P_F$ as a consequence of economy of scale behavior. The required $H$ also decreases but is still above the $H = 1$ limit even at $P_F = 3000$ MW. The value of $q_*$ decreases as $P_F$ increases but still remains safely above the $q_* = 3$ boundary. Overall, increasing the output power is desirable from many points of view except for the obvious one that very large plants may not be desirable by industry because of grid concerns, large capital investments, long construction times, etc.

Even more critical, when $P_F$ increases from 500 to 2500 MW the neutron wall loading increases from 1.7 to 8.5 MW/m$^2$ while the heat flux parameter increases from 339 to 1690 MW-T/m, both unacceptably large. In principle, for large power outputs, the driving constraints must be changed, as the design will now be driven more by technology than plasma physics [63].

The last sensitivity study involves the current drive efficiency $\eta_{CD}$. The fact that $\eta_{CD}$ is small even for the most efficient lower hybrid current system, would seem to imply that it plays an important role in the design of steady state reactors. If a more efficient method of current drive could be developed, would this lead to a more attractive design? To answer this question we have illustrated in Fig. 9 curves of $R_0$, $H$, $C_{MAG}$, $P_{VOL}$, $q_*$ versus $\eta_{CD}$. Observe that the trends are as expected, but somewhat surprisingly, not that large in magnitude. There is only a slight decrease in $R_0$ as $\eta_{CD}$ increases. Both $H$ and $C_{MAG}$ also decrease although not by much. Similarly, the corresponding increase in $P_{VOL}$ is small. The value of $q_*$ decreases because more current can now flow, but still remains safely above the $q_* = 3$ boundary. The reason for only modest changes is associated with the fact that the current profile for steady state reactors is assumed to be hollow. This leads to a large, perhaps uncomfortably large, bootstrap fraction, which then serves as the main contribution to the total plasma current. The current drive thus fortunately plays a smaller role so the low efficiency penalty is mitigated. The main conclusion is that if it is possible to stably maintain a hollow current profile, then one can live with low current drive efficiency.

In summary, our choice of parameters for the steady state reference reactor are reasonable in the context of sensitivity studies. The design is of reasonable size and cost. The two outstanding problems are the need for $H \approx 1.9$ and the still unresolved problem of divertor heat load.

- **Pulsed reference design**

The analysis for the pulsed reference reactor starts out similar to that of the steady state reactor. We apply our model allowing $B_C$ to float to see if a high TF field leads to the most desirable reactor in terms of smallest required $H$. Here perhaps a surprising result occurs. The maximum allowable TF field is not the best option. The reason is given shortly. Instead, the analysis leads to an alternate prescription for choosing the best value for $B_C$ from which it is then straightforward to evaluate the parameters for the pulsed reference reactor. The analysis is carried out assuming operation at the maximum Greenwald density limit $N_G = 0.85$. However, the analysis is also repeated for the minimum density option $N_G = 0.4$. The motivation is that in a pulsed machine it may be advantageous to trade off density versus temperature. High temperature has the positive effect of deceasing the plasma resistivity, thereby increasing the $L/R$ Ohmic decay time of the plasma. This should reduce the requirement on the OH transformer leading to a smaller reactor. Once the two reference reactors are designed, a set of sensitivity studies is carried out.

We begin by listing in Table 7.7 the input parameters for the high density option.

| Parameter | Symbol | Value |
|---|---|---|
| Greenwald density limit | $N_G$ | 0.85 |
| Elongation | $\kappa$ | 1.8 |
| Inverse aspect ratio | $\varepsilon$ | 0.25 |
| Beta limit | $\beta_N$ | 0.026 |
| Fusion gain | $Q$ | 26 |
| Fuel dilution factor | $f_D$ | 0.85 |
| Pulse length (hr) | $\tau_P$ | 1.5 |
| Maximum allowable TF stress (MPa) | $\sigma_{max}$ | 650 |
| Maximum allowable TF field (T) | $B_{max}$ | 23 |
| Maximum allowable OH stress (MPa) | $\hat{\sigma}_{max}$ | 500 |
| Maximum OH field (T) : $B_\Omega = \hat{B}_{max}$ | $\hat{B}_{max}$ | 23 |
| Thermal fusion power (MW) | $P_F$ | 500 |
| Blanket/shield thickness (m) | $b$ | 1 m |
| Number density profile factor | $\nu_n$ | 0.4 |
| Temperature profile factor | $\nu_T$ | 1.1 |
| Thermal conversion efficiency | $\eta_T$ | 0.4 |
| Wall to RF conversion efficiency for ICRH | $\eta_{RF}$ | 0.75 |

Table 7.7 Input parameters for the Pulsed reference design

Almost all of the input parameters for the pulsed and steady state reference reactors are the same in order to make a fair comparison. The differences involve replacing the steady state current drive efficiency $\eta_{CD}$ with equivalent pulsed parameters for the magnets: $\tau_P, \sigma_{max}, B_{max}, \hat{\sigma}_{max}, \hat{B}_{max}$. Also we have increased $\eta_{RF}$ from 0.5 to 0.75 to account for the higher efficiency of ICHR versus LHCD. In this connection we could reduce the required value of $Q$ from 26 to 17 as previously discussed to produce the same recirculating power. However, we have chosen to maintain $Q = 26$ to keep the physics comparison as fair as possible. Lastly, we have eliminated the current profile parameter $\nu_J$ as an input as it is actually an output for a pulsed reactor. The low density option uses the same parameters except that the density limit is reduced from $N_G = 0.85$ to $N_G = 0.4$.

The results from the scans in $B_C$ for the high and low density options are illustrated in Fig. 10. Keep in mind that in these scans the OH magnetic field has been set to $B_\Omega = \hat{B}_{max} = 23$ T as this leads to the smallest reactor. Observe the following points. Both high and low density options show the same trends. Favorably, the major radius $R_0$ decreases and the corresponding power density $P_{VOL}$ increases as the TF magnetic field increases over most of its interesting range, although $R_0$ only at a modest rate. On the other hand, unfavorably, both the required $H$ and cost $C_{MAG}$ increase with increasing $B_C$. This for many is a surprising result. The implication is that the "best" design makes use of the lowest possible TF field. This value of $B_C$ is determined by the requirement that $q_*$ remain above its kink limit $q_* = 3$ as shown in Fig. 10. Stated differently, the best value of $B_C$ corresponds to the situation where the Troyon $\beta$ limit and kink $q_*$ limit are both satisfied simultaneously.

The reason for this behavior is associated with the fact that a higher TF field requires more structure. Specifically, in the steady state case weak demands on the OH transformer allow room for the central hole size to shrink if more TF structure is added. In contrast, the TF structure for a pulsed reactor cannot be added that easily to fill in the central hole. That is, there is a penalty because the central hole size must be maintained to provide the required flux swing to generate the desired pulse length. Therefore, the additional structure needed for high field must be added in the outward direction, thereby increasing $R_0$. It thus competes with the anticipated reduction in $R_0$ due to improved plasma performance at higher $B_C$. This competition is why the $R_0$ curve is only a weakly decaying function of $B_C$ compared to the rapid decay for steady state.

With only a weakly changing $R_0$ we see intuitively that the cost, which is proportional to magnetic energy (i.e. $B_0^2 R_0^3$), should decrease as $B_C$ decreases. Next, at the beta limit, the necessity of maintaining a high power density (i.e. $\bar{n}_{20}^2 \bar{T}_k^2$) to provide the required fusion power, implies that $I_M \propto (\bar{n}_{20} \bar{T}_k) R_0 / B_0 \propto 1 / B_0$ must increase as $B_C$ decreases. This further implies that the kink safety factor $q_* \propto R_0 B_0 / I_M \propto B_0 / I_M$ should decrease with decreasing $B_C$. The increase in $I_M$ at low $B_C$ is also the reason why the total OH transformer radius, $R_\Omega + d$, increases – more flux is needed to drive a higher current. Finally, the required $H$ is determined by several competing effects, but is dominated by strong current dependence. Thus, the final conclusion is that $H$ must decrease as $B_C$ decreases as shown in Fig. 10.

The comparison of low versus high density also makes physical sense. At a fixed $B_C$, a lower density limit requires a higher temperature to maintain power density. The increased temperature does indeed lead to a smaller $R_0$ because of the longer $L/R$ time of the plasma. The reduced $R_0$ also leads to a lower cost $C_{MAG}$. In addition, at smaller $R_0$ the power density itself (i.e. $\bar{n}_{20}^2 \bar{T}_k^2$) must increase to produce the same required total fusion power ($P_F \propto \bar{n}_{20}^2 \bar{T}_k^2 R_0^3$). Now, the beta limit implies that an increasing power density requires a larger value of $I_M / R_0$ which in turn leads to a smaller value of $q_* \propto B_0 (R_0 / I_M)$. Again, the required $H$ involves a competition between several effects but is dominated by the strong $R_0$ dependence. The smaller $R_0$ leads to a larger required value of $H$.

The key conclusion from this discussion is that the constraint on $B_C$ for the "best" reactor is not $B_C = B_{max}$ as for steady state. Instead the requirement is that $B_C$ be chosen to make $q_* = 3$, the kink stability limit. A short calculation using Eq. (63) and the definition of $q_*$ shows that the $B_C$ constraint defining the reference pulsed reactor is given by,

|Pulsed|Steady State|
|---|---|

$$B_0(R_0) = \left(\frac{K_I}{K_q}\right)^{1/2} \frac{1}{R_0^{3/4}} \qquad\qquad B_C(R_0) = B_{\max}$$

$$B_C(R_0) = \frac{B_0}{1-\varepsilon_B} = \left(\frac{K_I}{K_q}\right)^{1/2} \frac{1}{(1-\varepsilon_B)R_0^{3/4}} \qquad\qquad B_0(R_0) = B_{\max}(1-\varepsilon_B) \tag{69}$$

In Eq. (69) the steady state constraint is also shown for comparison. The pulsed constraint is now used to close the system of equations defining the best pulsed reactor. A simple numerical calculation then yields the pulsed reference reactor design. Results from both the high density and low density options are presented in Table 7.8. Also shown for comparison are the relevant values for the steady state design.

There is a large amount of data which can be perused at leisure. Even so, several points stand out that are worth noting.

(a) All the designs require a value of $H$ that exceeds the limit $H = 1$. The pulsed reactors, however, require a smaller enhancement than the steady state reactor. The high density pulsed reactor requires $H = 1.35$ which is substantially lower than the steady state value of $H = 1.94$.

(b) The maximum TF magnetic field is only $B_C = 14.1$ T for the high $N_G$ pulsed reactor as compared to 23 T for the steady state reactor. The plasma magnetic field also has a comparable reduction from $B_0 = 11.6$ T to $B_0 = 7.6$ T.

(c) The major radius is somewhat larger for the pulsed system, $R_0 = 4.72$ m compared to $R_0 = 4.1$ m.

(d) Even so, the smaller $B_0$ dominates the larger $R_0$ resulting in a cost metric $C_{MAG}$ that is nearly a factor of two smaller for the pulsed reactor: 37.2 MJ/MW compared to 63.8 MJ/MW.

(e) As expected the current is higher in the pulsed reactor since the issue of low current drive efficiency is not important. The relevant currents are 7.89 MA compared to 5.53 MA.

(f) The lower TF field also leads to a lower, more desirable value for the divertor heat flux parameter with $h_\parallel$ reduced from 339 MW-T/m to 192 MW-T/m.

(g) In comparing the $N_G = 0.4$ low density pulsed reactor to the steady state reactor, we have a number of favorable results. Specifically, the TF field is smaller (17.7 : 23), the major radius is smaller (3.91 : 4.1), the cost is lower (32.7 : 63.8), the divertor heat flux is lower (266 : 339), and the required $H$ is smaller (1.80 : 1.94).

(h) The comparisons show that in many respects the low density pulsed reactor is more desirable than the high density pulsed reactor. The one counter point to this argument is that a higher value of $H$ is required (1.8 : 1.35) Thus, based on our definition of "best" reactor, the high density reactor is the best choice.

| Parameter | Symbol | Pulsed $N_G = 0.85$ | Pulsed $N_G = 0.4$ | Steady State |
|---|---|---|---|---|
| Kink safety factor | $q_*$ | 3 | 3 | 5.68 |
| Maximum field on the TF (T) | $B_C$ | 14.1 | 17.7 | 23 |
| Major radius (m) | $R_0$ | 4.72 | 3.91 | 4.10 |
| Minor radius (m) | $a$ | 1.18 | 0.979 | 1.02 |
| Plasma magnetic field (T) | $B_0$ | 7.60 | 8.75 | 11.6 |
| Plasma current (MA) | $I_M$ | 7.89 | 7.52 | 5.53 |
| Average density ($10^{20} \text{m}^{-3}$) | $\bar{n}_{20}$ | 1.53 | 1.00 | 1.43 |
| Average temperature (keV) | $\bar{T}_k$ | 9.11 | 18.5 | 12.1 |
| Confinement time (sec) | $\tau_E$ | 1.43 | 1.08 | 1.15 |
| Confinement enhancement factor | **H** | **1.35** | **1.80** | **1.94** |
| Bootstrap fraction | $f_B$ | 0.316 | 0.316 | 0.792 |
| On axis safety factor $q(\rho = 0)$ | $q_0$ | 1 | 1 | 4.75 |
| Current density profile factor | $\nu_J$ | 0.134 | 0.134 | (0.453) |
| TF magnet thickness (m) | $c$ | 0.351 | 0.455 | 0.874 |
| OH magnet thickness (m) | $d$ | 0.650 | 0.439 | --- |
| OH central hole size (m) | $R_\Omega$ | 1.54 | 1.05 | --- |
| Heating power/LH threshold | $(P_\alpha + P_A)/P_{LH}$ | 2.05 | 3.54 | 2.00 |
| Electric power out (MWe) | $P_E$ | 255 | 255 | 255 |
| ICRH wall power (MWe) | $P_{RF}$ | 25.6 | 25.6 | (38.5) |
| ICRH power absorbed (MW) | $P_A$ | 19.2 | 19.2 | (19.2) |
| Recirculating power fraction | $f_{RP}$ | 0.101 | 0.101 | 0.151 |
| Neutron wall loading (MW/m$^2$) | $P_W$ | 1.28 | 1.86 | 1.70 |
| Heat flux parameter (MW-T/m) | $h_\parallel$ | 192 | 266 | 339 |
| Stored TF magnetic energy (GJ) | $W_{TF}$ | 18.6 | 16.4 | 31.9 |
| Cost metric (MJ/MW) | $C$ | 37.2 | 32.7 | 63.8 |
| Power density metric (MW/m$^3$) | $P_{VOL}$ | 2.14 | 3.76 | 3.29 |

Table 7.8  Output parameters for the pulsed reference reactor

- **Pulsed sensitivity studies**

The interesting sensitivity parameters to examine for pulsed reactors are (a) the fusion power $P_F$, (b) the maximum allowable OH magnetic field $B_\Omega$, and (c) the required pulse length $\tau_P$. The inverse aspect ratio $\varepsilon$ is not as critical for a pulsed reactor as compared to a steady state reactor. Basically, smaller $\varepsilon$ allows for more room for the OH transformer which leads to the most desirable design, although the overall gains are modest. Below we present scans holding all input parameters fixed, including setting $q_* = 3$, and then varying, one by one, $P_F$, $B_\Omega$, and $\tau_P$.

The $P_F$ scan is illustrated in Fig. 11. As for the steady state reactor most quantities improve as the power output increases. Notably, the required $H$ decreases and actually reaches the value $H \approx 1$ for the $N_G = 0.85$ case when $P_F \approx 2.5 \text{ GW}$. The magnetic cost metric decreases while the power density increases. The maximum field at the TF coil $B_C$, increases with $P_F$, but still remains below the maximum allowable value of 23 T. Interestingly, the major radius $R_0$, and the total radius of the OH transformer $R_\Omega + d$ are nearly constant as $P_F$ increases. The increased compactness associated with a higher TF field competes with need for larger major radius for more output power, resulting in an $R_0$ which changes very little. Also, the increased $B_C$ leads to an increased $I_M$ at fixed $q_*$. This in turn competes with a higher temperature such as to keep $R_\Omega + d$ approximately constant.

Even assuming much higher output powers become acceptable to the US energy market, there are still the unresolved problems of neutron wall loading and heat flux to the divertor, which are substantially worse: when $P_F$ increases from 500 MW to 2500 MW, the neutron wall loading increases from 1.28 to 5.5 MW/m$^2$ while the heat flux parameter increases from 192 to 1260 MW-T/m. Both of these values are unacceptably large. At larger output powers the designs will be more limited by technology than plasma physics. As for steady state reactors, larger output powers require a new analysis. The design is now driven more by engineering constraints than plasma physics constraints [63].

Consider next the scan with $B_\Omega$ as shown in Fig. 12. The trends agree with intuition. A higher transformer field leads to a reduced major radius, a reduced transformer radius, a lower cost, and a higher power density. What is perhaps surprising is that changes are relatively modest. Indeed there is virtually no change in the required value of $H$. These results imply that access to a higher $B_\Omega$ helps, but does not appear to be a game changer for a pulsed reactor. The explanation is as follows. First, the radius of the OH transformer $R_\Omega + d$ makes only one contribution to the major radius. There are additional contributions due to the plasma radius $a$, the blanket/shield region $b$, and the TF coil thickness $c$. These other contributions dilute the effect of the transformer. Second, the flux $\psi_\Omega$ required to drive the desired flat top current remains approximately constant since its value is determined primarily by plasma physics. This flux scales as $\psi_\Omega \propto B_\Omega R_\Omega^2$. Thus for a constant $\psi_\Omega$ it follows that $R_\Omega \propto 1/B_\Omega^{1/2}$ has a weak dependence on $B_\Omega$. Third, even this $1/B_\Omega^{1/2}$ effect is diluted since increasing $B_\Omega$ leads to an increase in the transformer thickness $d \propto R_\Omega B_\Omega^2 \propto B_\Omega^{3/2}$. As the central hole size shrinks the magnet thickness increases. These three effects combine to produce only modest changes in $R_0$ and $R_\Omega + d$ as $B_\Omega$ increases.

The last scan of interest involves the pulse length $\tau_P$. The results are shown in Fig. 13. Here too, the trends are as expected. If the required pulse length becomes longer, this has unfavorable consequences. The major radius increases, the cost metric increases, and the power density decreases. These trends, however, are more modest than may have been anticipated. The reason is again associated with the fact that the OH transformer is only one contribution to $R_0$. In addition the OH transformer radius $R_\Omega$ and thickness $d$ are weak functions of $\tau_P$. Specifically, at fixed $B_\Omega$, the required flux swing is approximately equal to $\psi_\Omega = \pi B_\Omega R_\Omega^2 \approx R_2 I_\Omega \tau_P$ where $R_2$ is the secondary (i.e. plasma) resistance and $I_\Omega = I_M(1 - f_B)$ is the Ohmic contribution to the total plasma current. This implies that $R_\Omega \propto \tau_P^{1/2}$. Similarly, the thickness $d \propto R_\Omega B_\Omega^2 \propto \tau_P^{1/2}$. The overall transformer thickness thus has the weak scaling dependence $R_\Omega + d \propto \tau_P^{1/2}$. Because of dilution due to $a, b, c$ the $R_0$ dependence is even weaker.

## 8. Conclusions

The overall results have been presented and it is now time to draw major conclusions.

- Pulsed tokamak reactors should be reconsidered on the path forward to fusion energy. Compared to standard steady state reactors, pulsed reactors (corresponding to the lower $N_G$ option) are predicted to have comparable size, a lower cost TF magnet system, higher power density, and a smaller but comparable required enhancement of $H$.
- A key assumption in the analysis is the focus on lower power reactors, 500 MW versus the standard 2500 MW thermal fusion power. This choice is motivated by the current industrial view in the US that smaller, quicker to build, and lower capital cost reactors are more competitive. The design of small 500 MW reactors is driven largely by plasma physics constraints. In contrast, larger 2500 MW reactors require small or no enhancements in $H$ but are driven primarily by technology rather than plasma physics constraints. Heat load on the divertor and neutron wall loading are important technological constraints driving the design.
- All of the 500 MW reactors designed and discussed require enhancements in the value of $H$ above the standard empirical value of $H = 1$. Pulsed reactors require $H = 1.35$ while steady state reactors require $H = 1.9$. Accomplishing this goal will require advances in plasma physics. New modes of improved confinement have been discovered, but are not yet sufficiently robust and reliable to represent "standard" tokamak operation. Too much reliance on profile control would likely not be sufficiently robust. However, increasing the triangularity, which also appears to help confinement, may represent a good path forward. This is an important area for continuing plasma physics research.
- Access to higher magnetic fields is a potential game changer for steady state reactors. High field leads to reactors which are smaller, have higher power density, lower cost, and require the minimum enhancement in $H$. Setting $B_C = B_{max} \approx 23$ T is the best option.
- High field helps pulsed reactors but is not the same game changer as for steady state reactors. REBCO HTS tapes are still required for pulsed reactors since the optimum TF fields are typically on the order of 14 T – 17 T, above the capabilities of existing LTS $Nb_3Sn$ superconductors. In a pulsed reactor the value of the maximum TF field $B_C$ is determined by the requirement that the Troyon $\beta$ limit and kink current limit be satisfied simultaneously. One important consequence of a lower $B_C$ for pulsed reactors is that the TF cost metric is about 1/2 that of steady state reactors.

- The above points represent the main conclusions from our analysis. However, it is important to keep in mind that there is a family of "800 lb. gorillas" lurking in the tokamak living room that must be addressed before moving to fusion electricity. Common to both pulsed and steady state tokamaks are the need to improve $H$, solve the divertor heat flux and neutron wall loading problems, develop a way to survive major disruptions, and develop a workable blanket design, either solid or molten salt. New facilities will be needed to address these issues.
- Specific to steady state reactors is the need for robust sustainability of the hollow current density profile needed to maintain a high bootstrap fraction. This may become more difficult in the presence of large alpha heating.
- Specific to pulsed reactors is the need to develop large scale REBCO magnets. This seems realistic for the TF magnets, but the OH transformer is more difficult because of the relatively rapidly varying flux swings. Also, there is a relatively high uncertainty about the number of possible cycles and OH transformer replacement time, which has a direct impact on the required $\tau_P$. These issues can and should be addressed in small scale D-D facilities.

Overall, the tokamak appears to be the fastest way forward to fusion electricity in terms of a plausibly sized reactor with high power density and reasonable costs. However, the problems that remain indicate that the research phase of fusion is not yet complete either in plasma physics or fusion technology.

**Acknowledgements**

There have been many, many, people who have helped us in the course of our work, too many to actually list. Nevertheless, we would like to acknowledge a few MIT colleagues for their substantial contributions to our understanding of the issues: Paul Bonoli, Leslie Bromberg, Bob Granetz, Martin Greenwald, Ian Hutchinson, Brian Labombard, Joe Minervini, Bob Mumgaard, Miklos Porkolab, Brandon Sorbom, Anne White, and Dennis Whyte.

AJC was partially supported by the Simons Foundation/SFARI (560651,AB), and the US DoE, Fusion Energy Sciences under Award Nos. DE-FG02-86ER53223 and DE-SC0012398. JPF was supported by DoE, Fusion Energy Sciences under Award No. DE-FG02-91ER54109.

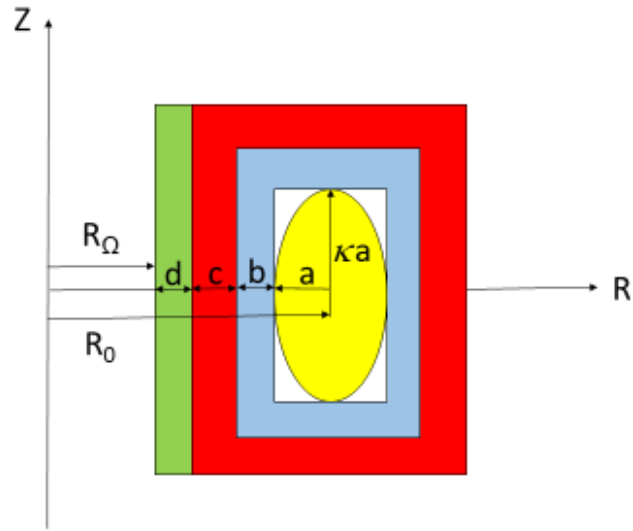

Figure 1

Simplified tokamak geometry valid for both steady state and pulsed reactors

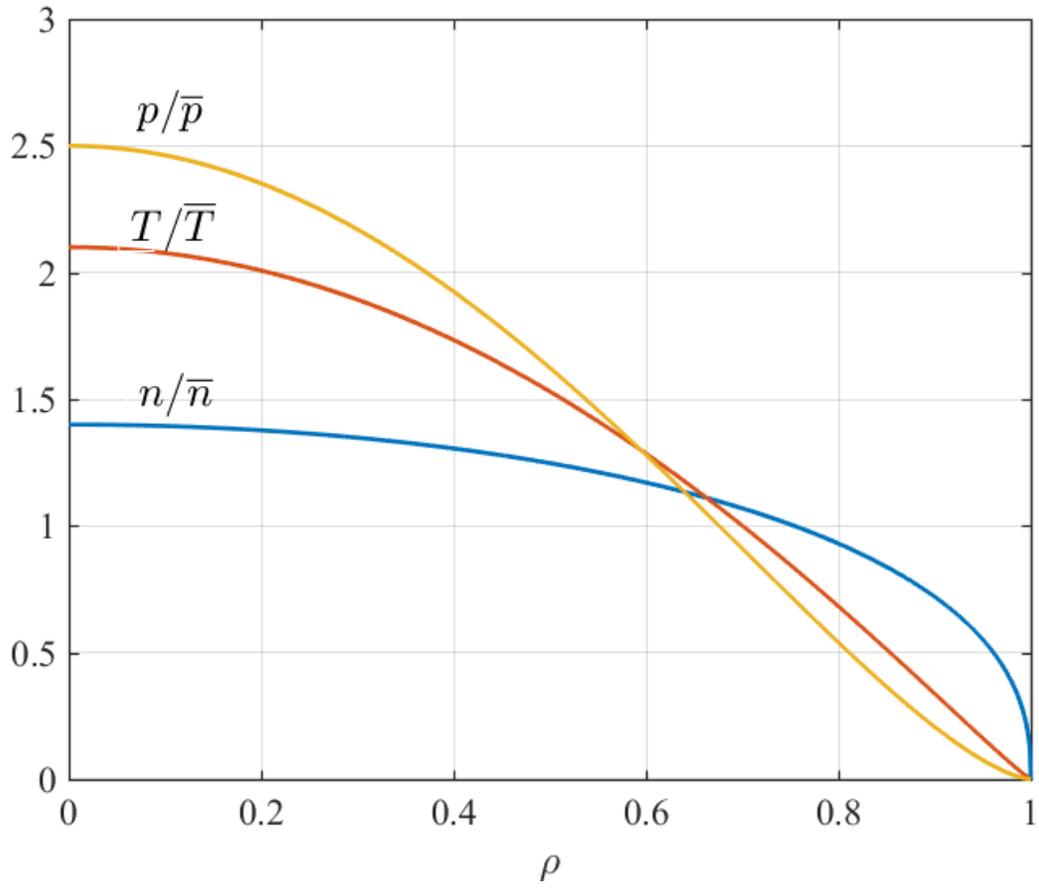

Figure 2a

Curves of density $n$, temperature $T$, and pressure $p$ versus the normalized flux radius $\rho$ for $\nu_n = 1.4$ and $\nu_T = 1.1$

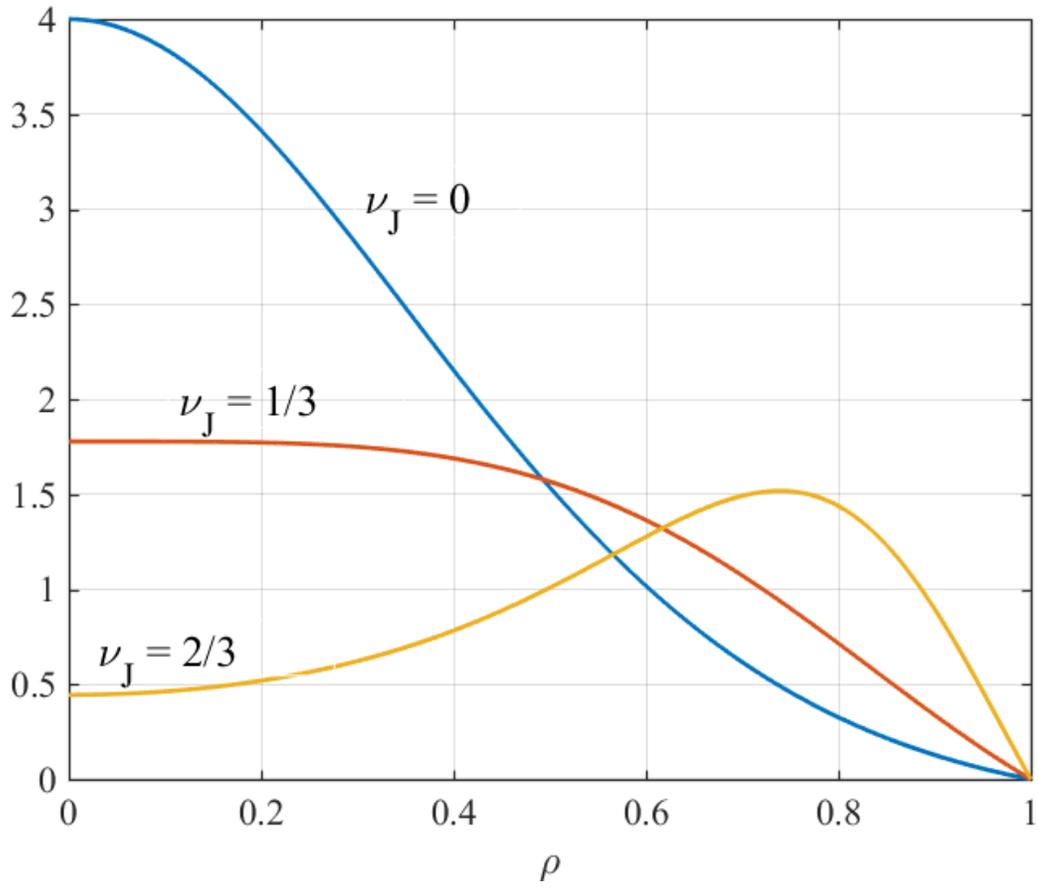

Figure 2b

Curves of the toroidal current density $J_\phi$ versus normalized flux radius $\rho$ for several values of the current profile parameter $\nu_J$

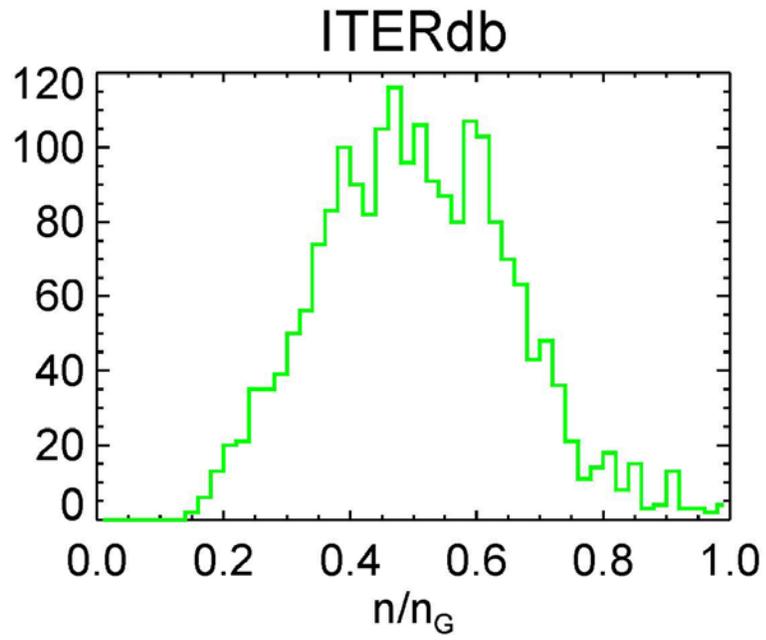

Figure 3

Number of experimental shots entering H-mode as obtained from the ITER data-base

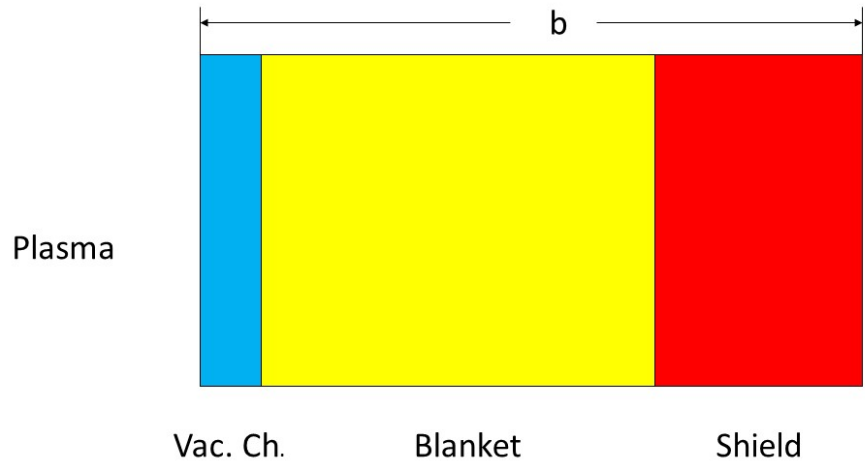

Figure 4

The blanket region consisting of the vacuum chamber, blanket, and shield. Note that the plasma is on the left.

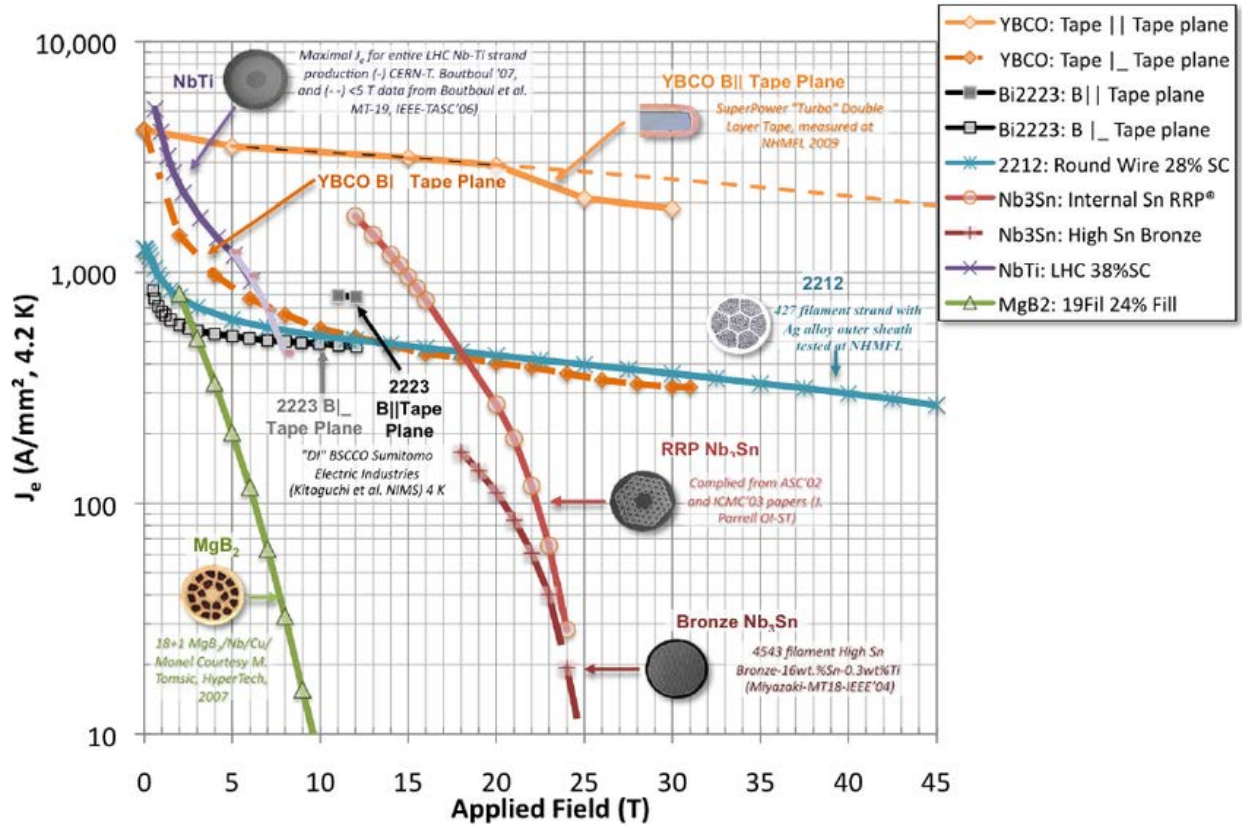

Figure 5

Plot of critical current density of various superconductors versus applied magnetic field at $T \approx 4$ °K

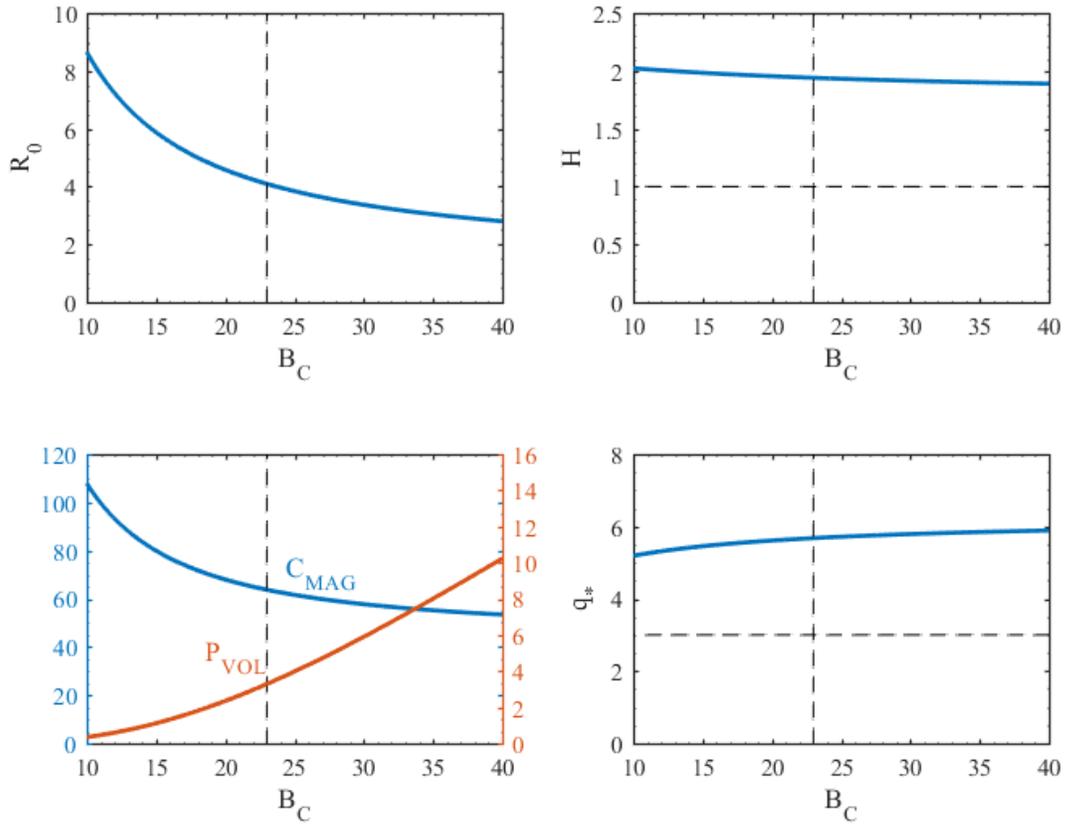

Figure 6

Critical parameters $R_0(\mathrm{m})$, $H$, $C_{MAG}(\mathrm{MJ/MW})$, $P_{VOL}(\mathrm{MW/m^3})$, and $q_*$ versus $B_C(\mathrm{T})$ for the steady state reference reactor

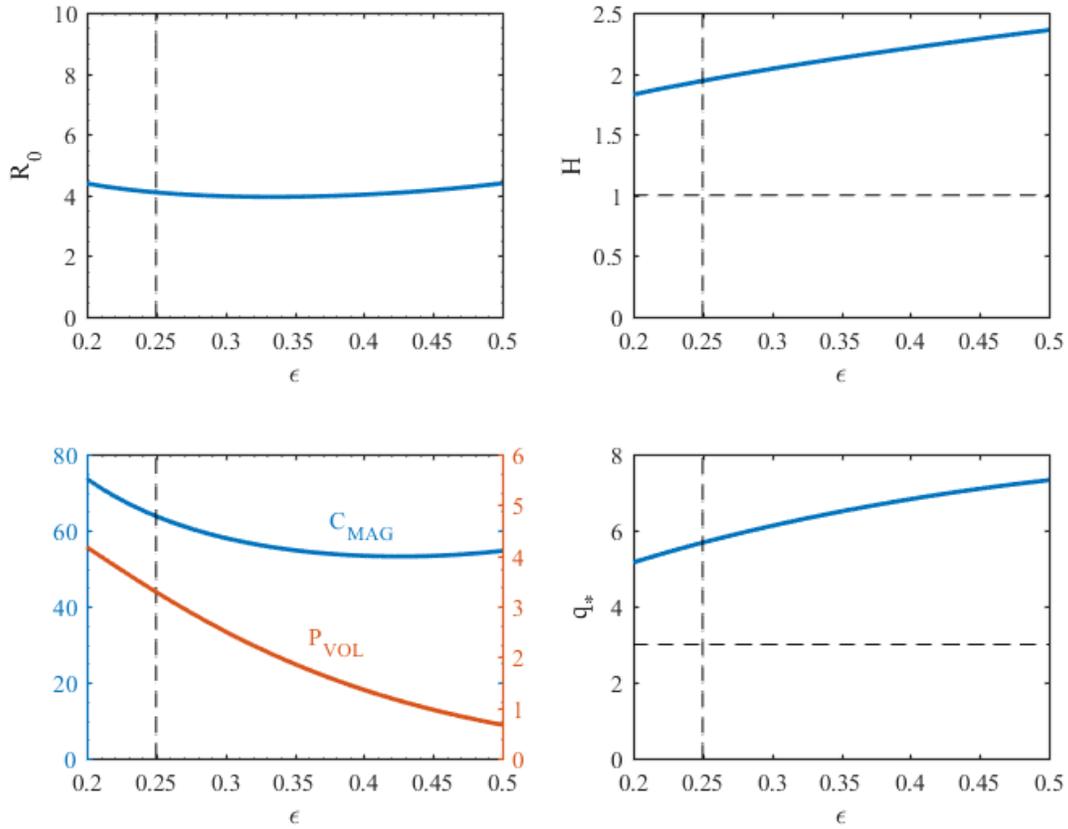

Figure 7

Critical parameters $R_0(\mathrm{m})$, $H$, $C_{MAG}(\mathrm{MJ/MW})$, $P_{VOL}(\mathrm{MW/m^3})$, and $q_*$ versus $\varepsilon$ for the steady state reference reactor

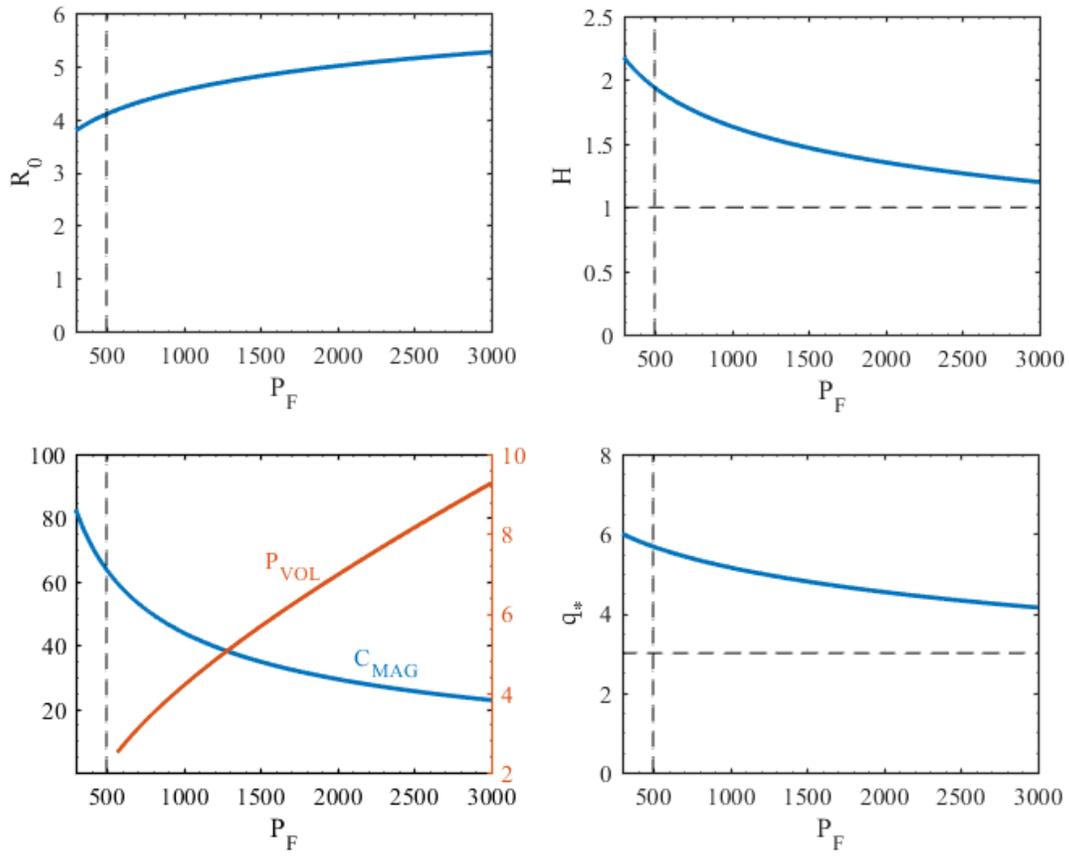

Figure 8

Critical parameters $R_0(\text{m})$, $H$, $C_{MAG}(\text{MJ/MW})$, $P_{VOL}(\text{MW/m}^3)$ and $q_*$ versus $P_F(\text{MW})$ for the steady state reference reactor

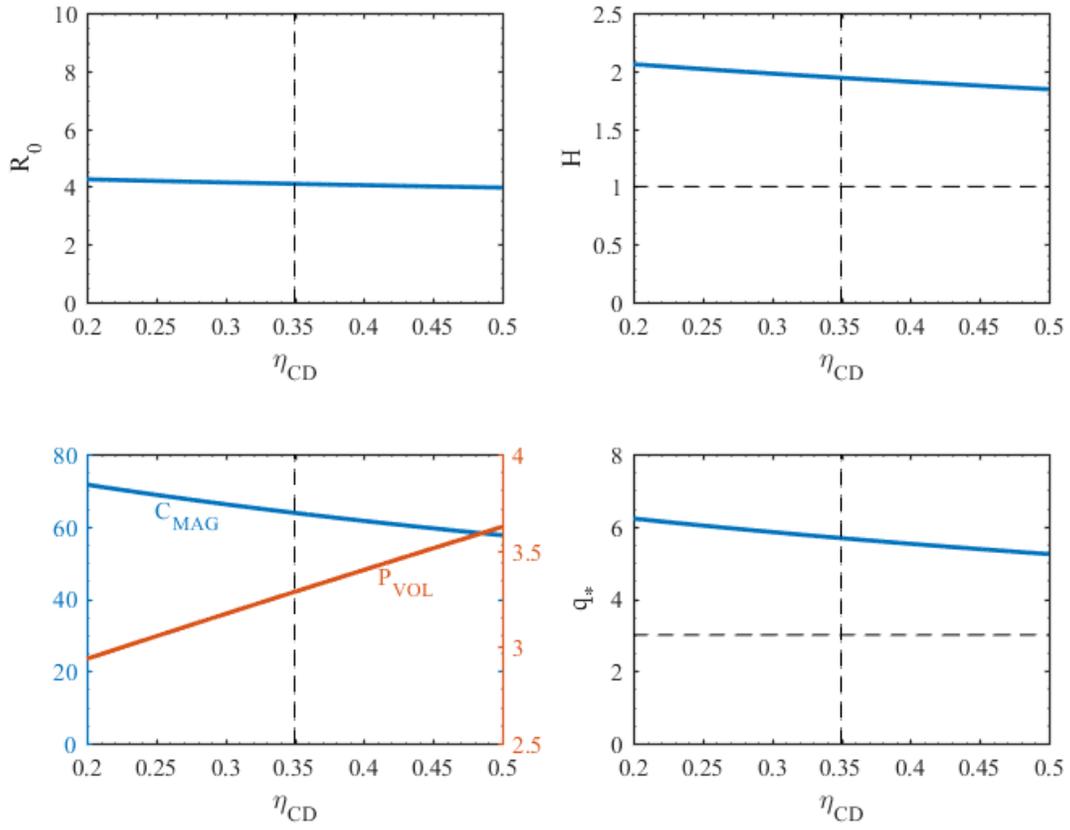

Figure 9

Critical parameters $R_0$(m), $H$, $C_{MAG}$(MJ/MW), $P_{VOL}$(MW/m$^3$), and $q_*$ versus $\eta_{CD}$ for the steady state reference reactor

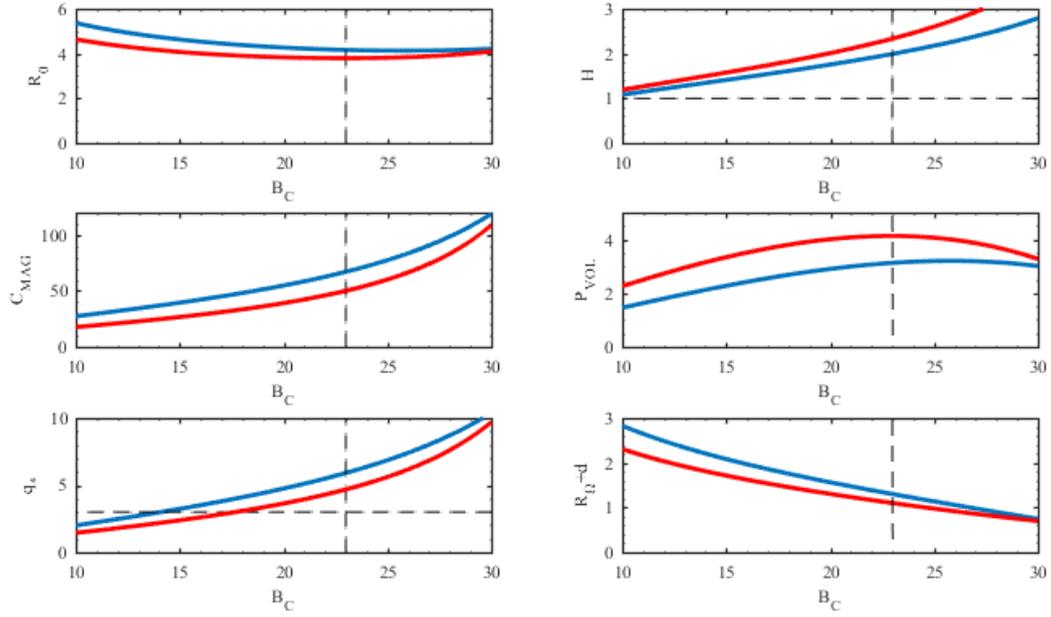

Figure 10

Critical parameters $R_0$ (m), $H$, $C_{MAG}$ (MJ/MW), $P_{VOL}$ (MW/m$^3$), $q_*$, and $R_\Omega + d$ (m) versus $B_C$ for the pulsed reference reactor. The $N_G = 0.85$ high density option is in blue while the $N_G = 0.4$ low density option is in red.

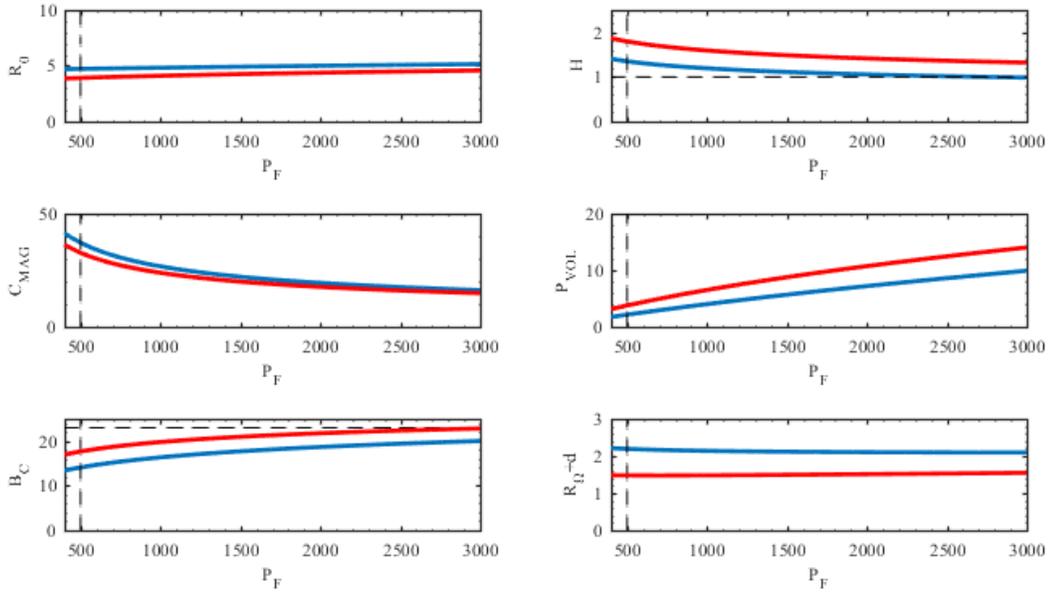

Figure 11

Critical parameters $R_0$(m), $H$, $C_{MAG}$(MJ/MW), $P_{VOL}$(MW/m$^3$), $B_C$(T), and $R_\Omega + d$ (m) versus $P_F$ for the pulsed reference reactor. The $N_G = 0.85$ high density option is in blue while the $N_G = 0.4$ low density option is in red.

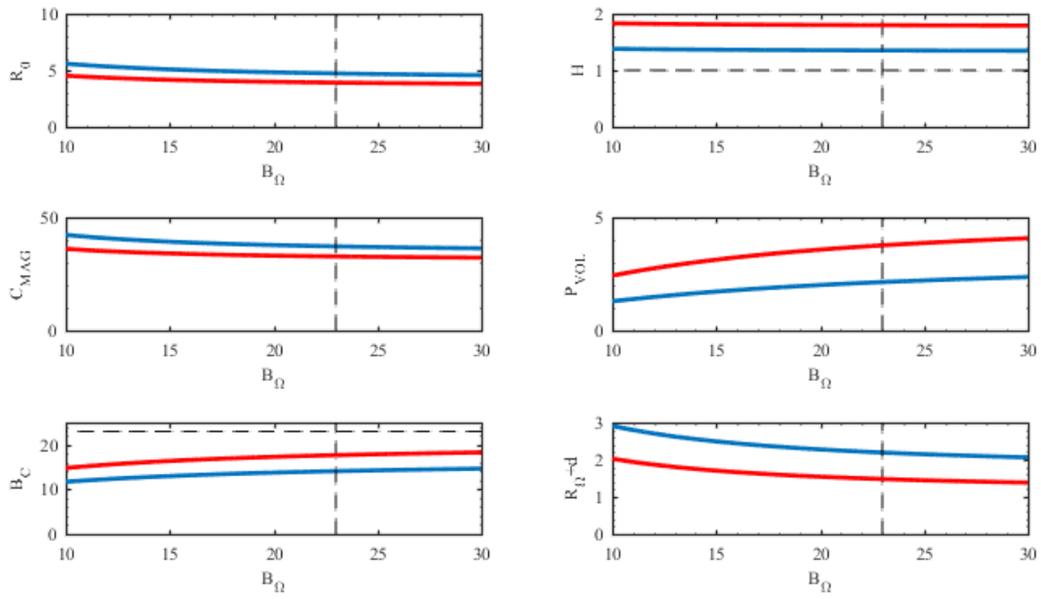

Figure 12

Critical parameters $R_0$(m), $H$, $C_{MAG}$(MJ/MW), $P_{VOL}$(MW/m$^3$), $B_C$(T), and $R_\Omega + d$ (m) versus $B_\Omega$ for the pulsed reference reactor. The $N_G = 0.85$ high density option is in blue while the $N_G = 0.4$ low density option is in red.

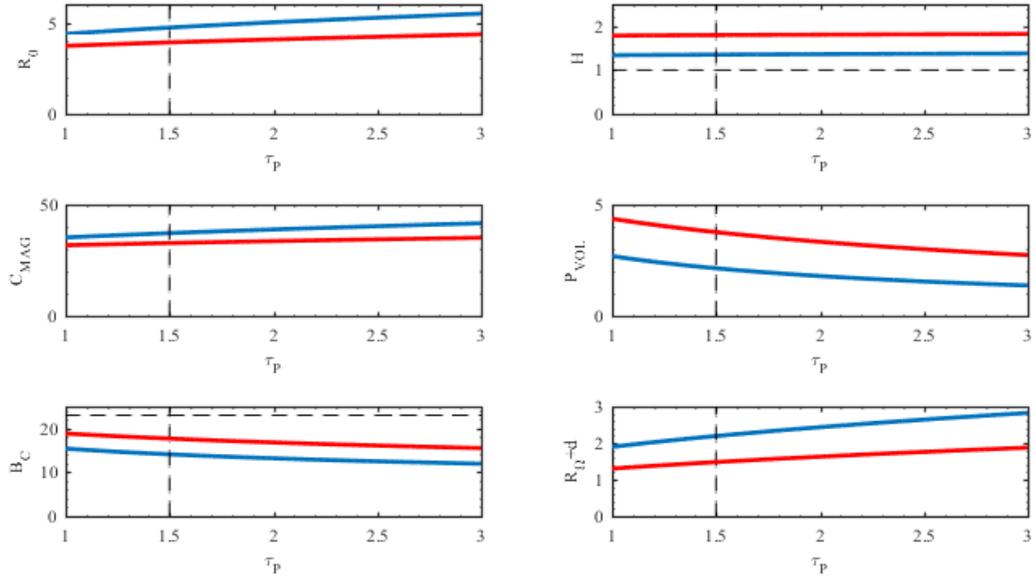

Figure 13

Critical parameters $R_0$(m), $H$, $C_{MAG}$(MJ/MW), $P_{VOL}$(MW/m$^3$), $B_C$(T), and $R_\Omega + d$ (m) versus $\tau_P$ for the pulsed reference reactor. The $N_G = 0.85$ high density option is in blue while the $N_G = 0.4$ low density option is in red.

# Appendix A
# The Bootstrap Current Fraction

The starting point for the analysis is the general expression for the bootstrap current in a tokamak with arbitrary cross section [A1]. This expression is simplified by assuming (1) equal temperature electrons and ions $T_e = T_i = T$, (2) large aspect ratio $\varepsilon \ll 1$, and (3) negligible collisionality $\nu_* \to 0$. The bootstrap current $\mathbf{J}_B \approx J_B \mathbf{e}_\phi$, with derivatives expressed in terms of the poloidal flux $\psi$, reduces to

$$J_B = -3.32 f_T R_0 n T \left[ \frac{1}{n}\frac{dn}{d\psi} + 0.054 \frac{1}{T}\frac{dT}{d\psi} \right] \tag{A1}$$

A slightly optimistic approximate form [A2] for the trapped particle fraction $f_T$ that makes use of the elliptic flux surface model is given by

$$f_T(\psi) \approx 1.46\left[1 - \frac{B_{\min}(\psi)}{B_{\max}(\psi)}\right]^{1/2} \approx 1.46\left[1 - \frac{R_{\min}(\psi)}{R_{\max}(\psi)}\right]^{1/2} \approx 1.46\varepsilon^{1/2}\rho^{1/2} \tag{A2}$$

Here, as in the main text, $\rho$ is a radial-like flux surface label that varies between $0 \leq \rho \leq 1$. In other words $\psi = \psi(\rho)$. Under these assumptions the bootstrap current for our density and temperature profiles can be written (in practical units) as

$$\begin{aligned}J_B(x) &= -0.07764\,\varepsilon^{1/2} \frac{R_0 n_{20} T_k x^{1/4}}{d\psi/dx}\left[\frac{1}{n_{20}}\frac{dn_{20}}{dx} + 0.054\frac{1}{T_k}\frac{dT_k}{dx}\right] \\ &= 0.07764(1+\nu_n)(1+\nu_T)(\nu_n + 0.054\nu_T)\varepsilon^{1/2} R_0 \bar{n}_{20}\bar{T}_k \frac{x^{1/4}(1-x)^{\nu_p - 1}}{d\psi/dx} \text{ MA/m}^2 \end{aligned} \tag{A3}$$

Here, $x = \rho^2$ and $\nu_p = \nu_n + \nu_T$.

- **Evaluation of** $d\psi/dx$

What remains for the evaluation of $J_B(x)$ is the calculation of $\psi' \equiv d\psi/dx$. Keep in mind that at this point, in spite of the approximations that have been made, the expression for $J_B(x)$ is still valid for arbitrary cross section.

The analysis that follows shows how to calculate $\psi'$ in terms of the normalized overall current density profile $j(x)$. The analysis makes use of Ampere's law, plus the concentric ellipse model for the flux surfaces. Ampere's law applied over a given elliptic flux surface $x = \text{constant}$ is given by

$$\oint \mathbf{B}_P \cdot d\mathbf{l} = \mu_0 \hat{I}(x) \tag{A4}$$

Consider first the right hand side. In this expression the plasma current (in MA) flowing within the given flux surface can be expressed as

$$\begin{aligned}\hat{I}(x) &= -\int J_P dS = -\pi a^2 \kappa \int_0^x \overline{J}_P(x)\,dx = I_M \int_0^x j(x)\,dx \quad \text{MA} \\ \overline{J}_P(x) &= \frac{1}{2\pi}\int_0^{2\pi} d\alpha\, J_P(x,\alpha) \quad \text{MA/m}^2\end{aligned} \tag{A5}$$

with the normalized current density defined as

$$j(x) = -\frac{\overline{J}_P(x)}{I_M/\pi a^2 \kappa} \qquad \int_0^1 j(x)\,dx = 1 \tag{A6}$$

The normalization constraint is a consequence of the requirement $\hat{I}_M(1) = I_M$.

Turning to the left hand side of Eq. (A4), we note that for the elliptic flux surface model, (a) $\psi(x,\alpha) \to \psi(x)$ and (b) $x = \text{constant}$ on a given flux surface, (implying that $dx = 0$). It then follows that

$$\mathbf{B}_P = \frac{1}{R}\nabla\psi \times \mathbf{e}_\phi = \frac{1}{R}\left(\psi_R \mathbf{e}_Z - \psi_Z \mathbf{e}_R\right) \approx \frac{1}{R_0}\frac{d\psi}{dx}\left(x_R \mathbf{e}_Z - x_Z \mathbf{e}_R\right)$$

$$d\mathbf{l} = dR\mathbf{e}_R + dZ\mathbf{e}_Z = \left(R_\alpha \mathbf{e}_R + Z_\alpha \mathbf{e}_Z\right)d\alpha \tag{A7}$$

$$\oint \mathbf{B}_P \cdot d\mathbf{l} = \frac{1}{R_0}\frac{d\psi}{dx}\int_0^{2\pi}\left(x_R Z_\alpha - x_Z R_\alpha\right)d\alpha$$

The derivatives in the integral in the last equation can be easily evaluated using the elliptic flux surface representation

$$\left.\begin{array}{l} R = R_0 + ax^{1/2}\cos\alpha \\ Z = \kappa a x^{1/2}\sin\alpha \end{array}\right\} \quad \rightarrow \quad \frac{(R-R_0)^2}{a^2} + \frac{Z^2}{\kappa^2 a^2} = x(\psi)$$

$$R_\alpha = -ax^{1/2}\sin\alpha \qquad x_R = \frac{2}{a}x^{1/2}\cos\alpha \tag{A8}$$

$$Z_\alpha = \kappa a x^{1/2}\cos\alpha \qquad x_Z = \frac{2}{\kappa a}x^{1/2}\sin\alpha$$

Using these derivatives and carrying out the $\alpha$ integration leads to

$$\oint \mathbf{B}_P \cdot d\mathbf{l} = 2\pi\left(\frac{1+\kappa^2}{\kappa}\right)\frac{x}{R_0}\frac{d\psi}{dx} \tag{A9}$$

Equating Eq. (A9) and (A5) yields the required expression (in practical units) for $\psi'$,

$$x\frac{d\psi}{dx} = \frac{1}{5}\frac{\kappa}{1+\kappa^2}I_M R_0 \int_0^x j(x)\,dx \tag{A10}$$

After substituting this expression into Eq. (A3) we obtain the general expression for the normalized bootstrap current

$$j_B(x) = \frac{J_B(x)}{I_M / \pi a^2 \kappa}$$

$$= 2.440(1+\nu_n)(1+\nu_T)(\nu_n + 0.054\nu_T)\varepsilon^{5/2}\kappa^{1.27} \frac{R_0^2 \bar{n}_{20} \bar{T}_k}{I_M^2} \frac{x^{5/4}(1-x)^{\nu_p - 1}}{\int_0^x j(x)\,dx} \quad (A11)$$

For our specific current profile, repeated here for convenience,

$$j(x) = 4(1-\nu_J)^2 \frac{(1-x)(1+3\nu_J x)}{[1+(1-3\nu_J)x + \nu_J x^2]^3} \quad (A12)$$

we can analytically evaluate the integral in the denominator of Eq. (A11).

$$\int_0^x j(x)\,dx = 4(1-\nu_J)^2 \frac{x}{[1+(1-3\nu_J)x + \nu_J x^2]^2} \quad (A13)$$

The expression for the bootstrap current density reduces to

$$j_B(x) = 0.6099(1+\nu_n)(1+\nu_T)(\nu_n + 0.054\nu_T)\varepsilon^{5/2}\kappa^{1.27}\frac{R_0^2 \bar{n}_{20}\bar{T}_k}{I_M^2}\hat{j}_B(x)$$
$$\hat{j}_B(x) = \frac{1}{(1-\nu_J)^2} x^{1/4}(1-x)^{\nu_p - 1}[1+(1-3\nu_J)x + \nu_J x^2]^2 \quad (A14)$$

- **The bootstrap fraction**

The expression for $j_B(x)$ can now be integrated over the plasma cross section to yield the bootstrap fraction $f_B$. A straightforward, slightly tedious, calculation leads to the complicated but analytic expression

$$f_B = \frac{I_B}{I} = \frac{\int_0^1 j_B(x)\,dx}{\int_0^1 j(x)\,dx} = \int_0^1 j_B(x)\,dx = K_B \frac{\bar{n}_{20}\bar{T}_k R_0^2}{I_M^2}$$

$$K_B = 0.6099(1+\nu_n)(1+\nu_T)(\nu_n + .054\nu_T)\kappa^{1.27}\varepsilon^{5/2}\, C_B(\nu_J,\nu_p)$$

$$C_B = \frac{1}{(1-\nu_J)^2}\int_0^1 x^{1/4}(1-x)^{\nu_p - 1}\left[1+(1-3\nu_J)x + \nu_J x^2\right]^2 dx$$

$$= \frac{\Gamma(\nu_p)}{(1-\nu_J)^2}\left[\frac{\Gamma(5/4)}{\Gamma(\nu_p + 5/4)} + 2(1-3\nu_J)\frac{\Gamma(9/4)}{\Gamma(\nu_p + 9/4)} + (1 - 4\nu_J + 9\nu_J^2)\frac{\Gamma(13/4)}{\Gamma(\nu_p + 13/4)}\right.$$

$$\left. + 2\nu_J(1-3\nu_J)\frac{\Gamma(17/4)}{\Gamma(\nu_p + 17/4)} + \nu_J^2 \frac{\Gamma(21/4)}{\Gamma(\nu_p + 21/4)}\right]$$

(A15)

The function $C_B(\nu_J,\nu_p)$ is illustrated in Fig. A1 for $\nu_p = 1.5$.

- **Choosing $\nu_J$**

We see from Fig. A1 that the bootstrap fraction has a strong dependence on $\nu_J$. The question that naturally arises is "how do we choose $\nu_J$?". Typical values are quite different for steady state and pulsed devices.

For a steady state device, the total current is relatively small because of current drive inefficiency and the need for high fusion gain. We assume that current drive is primarily produced by lower hybrid waves (LHCD), which, in general, produce a profile with an off axis maximum, designed to overlap as much as possible with the bootstrap current density. Thus, the bootstrap and LHCD current density profiles are both hollow. A small amount of current drive may be added by ion cyclotron waves (ICCD) to fill in the profile near the axis. The corresponding ICCD power is relatively small and for simplicity, and slightly optimistically, is assumed to generate current with the same $\eta_{CD}$ as LHCD.

These observations suggest that a plausible criterion for determining the steady state $\nu_J$ is to require that the profiles for the total current and the fractional bootstrap current have an off-axis maximum at the identical radius. Clearly, this implies that the current drive profile will also have a maximum at this same radius. Mathematically, the criterion for determining $\nu_J$ can be written as

$$\left.\frac{d}{dx}j(x,\nu_J)\right|_{x=x_{max}} = \left.\frac{d}{dx}j_B(x,\nu_J)\right|_{x=x_{max}} = 0 \quad \text{(A16)}$$

Note that the resulting $\nu_J$ is only a function of $\nu_p$. The criterion is easy to evaluate numerically and a good approximation for the solution in the practical range $1.4 < \nu_p < 2$ is

$$\nu_J \approx 0.453 - 0.1(\nu_p - 1.5) \quad \text{(A17)}$$

For $\nu_p = 1.5$ the off-axis peak is located at $\rho_{max} = x_{max}^{1/2} = 0.518$.

Consider next pulsed reactors. No current drive is required and heating is produced by alpha particles plus on-axis ion cyclotron heating (ICH). The current tends to be as large as possible to maximize confinement, but is subject to both the kink and beta instability constraints. The bootstrap fraction is smaller than in steady state reactors. Note that the total current density profile is peaked on axis.

The pulsed reactor value of $\nu_J$ is determined by simultaneously satisfying two constraints. First, to avoid kink driven disruptions we require $q_* > 2.5$. Second, because of the relatively large current we assume the plasma will operate in the sawtooth regime, implying that $q(0) \equiv q_0 \approx 1$. The value of $\nu_J$ can now be easily calculated.

The kink safety factor is simply a definition given by

$$q_* = \frac{2\pi a^2 B_0}{\mu_0 R_0 I}\left(\frac{1+\kappa^2}{2}\right) = \frac{5a^2 B_0}{R_0 I_M}\left(\frac{1+\kappa^2}{2}\right) \quad \text{(A18)}$$

The local safety factor can be expressed as

$$q(x) = \frac{F(x)}{2\pi} \int \frac{dl}{R^2 B_p} \approx \frac{B_0}{2\pi \psi'} \int_0^{2\pi} \left( \frac{R_\alpha^2 + Z_\alpha^2}{x_R^2 + x_Z^2} \right)^{1/2} d\alpha = \frac{B_0 a^2 \kappa}{2\psi'(x)} \tag{A19}$$

with $F(x) = RB_\phi \approx R_0 B_0$. We next eliminate $\psi'$ by means of Eq. (A10), leading to

$$q(x) = q_* \frac{x}{\int_0^x j(x)\,dx} = q_* \frac{[1 + (1 - 3\nu_J)x + \nu_J x^2]^2}{4(1 - \nu_J)^2} \tag{A20}$$

The pulsed reactor value of $\nu_J$ is determined by setting $x = 0$ and $q(0) \equiv q_0 \approx 1$. We obtain

$$\nu_J = 1 - \left( \frac{q_*}{4q_0} \right)^{1/2} = 0.209 \tag{A21}$$

The numerical value corresponds to $q_* = 2.5$.

To summarize, the bootstrap fraction is given by Eq. (A15) with the steady state and pulsed reactor values of $\nu_J$ given by Eqs. (A17) and (A21) respectively.

**References**

[A1] P. Helander and D. J. Sigmar, Collisional Transport in Magnetized Plasmas (Cambridge University Press, Cambridge, England, 2002)

[A2] Y. R. Lin-Liu and R. L. Miller, Physics of Plasmas **2**, 1666 (1995)

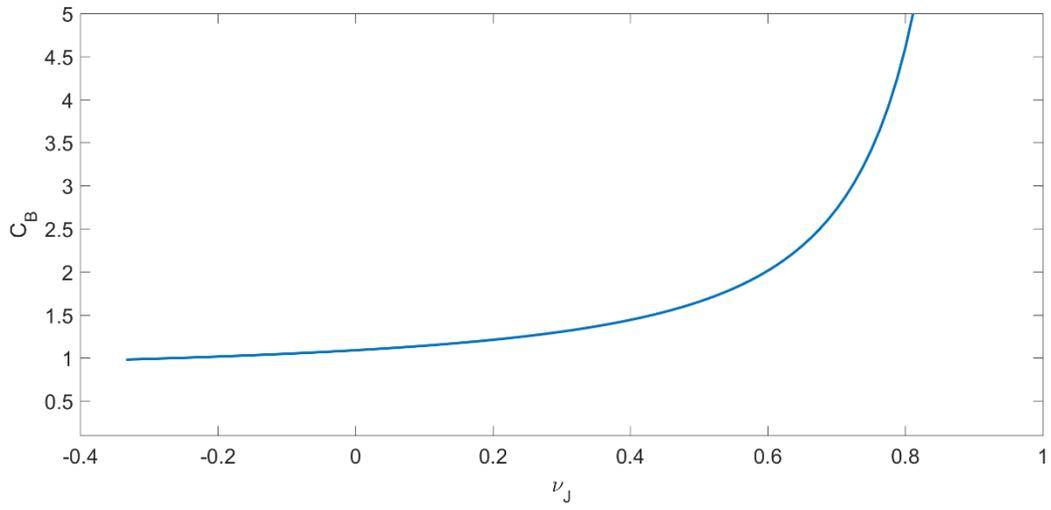

Figure A1

Illustration of $C_B$ versus $\nu_J$ for $\nu_p = 1.5$

# Appendix B

# The Toroidal Field Magnet

The goal of Appendix B is to determine the dimensions of the toroidal field (TF) magnet as a function of $R_0, B_0$. This is particularly important in the design of a pulsed tokamak reactor.

## B.1 The overall magnet thickness

The magnet model assumes that the total coil thickness $c$ is comprised of four contributions

$$c = c_S + c_J + c_{CU} + c_{He} \tag{B1}$$

Here, $c_S$ is the thickness of structural material to mechanically support the magnet stresses, $c_J$ is the thickness of the superconducting winding stack needed to carry the TF coil current, $c_{CU}$ is the thickness of copper to prevent overheating in case of a partial or full quench, and $c_{He}$ is the equivalent thickness of helium coolant to keep the magnet superconducting. As we shall see, most of the TF magnet thickness is due to the structural support material. The thicknesses are calculated as follows.

## B.2 Magnet forces and stresses

Consider first $c_S$. The three forces that contribute to the TF magnet stress are (1) the hoop force, (2) the centering force, and (3) the out of plane bending force. The largest contributions arise from the centering and tensile forces and thus for simplicity, the bending force is neglected. The strategy is to separately calculate the stresses due to the tensile hoop and compressional centering forces. These are combined to form the Tresca stress, which is then set equal to the maximum allowable stress. Use of the Tresca stress is made for convenience since it leads to a simple analytic expression for $c_S$. Keep in mind that the maximum allowable average material stress $\sigma_{max}$ for a high strength cryogenic structural material such as Inconel 718 is on the order of $600-700$ MPa.

For simplicity we assume the magnet structure is monolithic when calculating forces; that is, the magnet is a single, axisymmetric, structure, with no gaps between separate coils. This is a reasonable approximation for calculating forces. However, when calculating stresses the discrete structure of each coil must be, and is, included.

- **The tensile hoop force and stress**

The quantity $c_S$ is calculated as follows. For the tensile force, we split the TF magnet into an upper and lower half as shown in Fig. B1 and calculate the upward force on the top half of the magnet due to the magnetic field. The result is

$$F_Z = \int \mathbf{J} \times \mathbf{B} \cdot \mathbf{e}_Z \, d\mathbf{r} = \frac{1}{\mu_0} \int \left[ \mathbf{B} \cdot \nabla \mathbf{B} - \nabla \frac{B^2}{2} \right] \cdot \mathbf{e}_Z \, d\mathbf{r}$$
$$= -\frac{1}{2\mu_0} \int \nabla \cdot \left( B^2 \mathbf{e}_Z \right) d\mathbf{r} = -\frac{1}{2\mu_0} \int_{S_{in}} \left( B_\phi^2 \mathbf{n} \cdot \mathbf{e}_Z \right) dS \quad \text{(B2)}$$

Here, for the TF magnet we have written $\mathbf{B} = B_\phi(R,Z) \mathbf{e}_\phi$. Also, $S_{in}$ denotes the inner surface of the magnet with $\mathbf{n}$ the corresponding outward normal (which actually points in the inward direction towards $R=0$). The contribution from the outer surface $S_{out}$ vanishes because $B_\phi$ is zero on this surface.

Now, for an arbitrary shaped TF magnet whose inner surface is parameterized as $R = R(\theta), Z = Z(\theta)$, it follows that

$$dS = 2\pi R(R_\theta^2 + Z_\theta^2)^{1/2} d\theta$$

$$\mathbf{n} = \frac{-Z_\theta \mathbf{e}_R + R_\theta \mathbf{e}_Z}{(R_\theta^2 + Z_\theta^2)^{1/2}} \tag{B3}$$

With $B_\phi$ on the surface given by $B_\phi = B_0 R_0 / R$, we see that Eq. (B2) reduces to

$$F_Z = -\frac{\pi R_0^2 B_0^2}{\mu_0} \int_0^\pi \frac{R_\theta}{R} d\theta = \frac{\pi R_0^2 B_0^2}{\mu_0} \ln\left(\frac{1+\varepsilon_B}{1-\varepsilon_B}\right) \tag{B4}$$

where $\varepsilon_B = (a+b)/R_0$. Interestingly, the upward force is independent of the magnet shape, current density profile, and thickness.

Next, note that $F_Z$ is balanced by the two tensile forces $F_{T1}$ and $F_{T2}$ at the bottom faces of the upper half of the TF magnet. In other words $F_Z = F_{T1} + F_{T2}$. For a magnet with approximately constant tension around its perimeter, then $F_{T1} = F_{T2} = F_T$ and $F_Z = 2F_T$.

Using the assumption that neighboring coils are in wedging contact with each other on the inboard side it is then easy to calculate the total inboard tensile force produced TF magnets. We find

$$F_T = \sigma_T A_T = \pi \sigma_T \left[(R_0 - a - b)^2 - (R_0 - a - b - c_S)^2\right] \approx 2\pi \sigma_T c_S R_0 (1 - \varepsilon_B) \tag{B5}$$

Here, $\sigma_T$ is the portion of the maximum allowable stress $\sigma_{\max}$ that balances the tensile forces. Also, for analytic simplicity we have made the reasonable thin coil approximation $c_S \ll 2R_0(1 - \varepsilon_B)$. Setting $F_Z = 2F_T$ leads to an expression for $\sigma_T$,

$$\sigma_T = \frac{B_0^2 R_0}{4\mu_0 c_S} \left[\frac{1}{(1-\varepsilon_B)} \ln\left(\frac{1+\varepsilon_B}{1-\varepsilon_B}\right)\right] \tag{B6}$$

- **The centering magnetic force and stress**

A similar analysis holds for the centering force, which following the derivation in Eq. (B2), can be written as

$$F_R = \int \mathbf{J} \times \mathbf{B} \cdot \mathbf{e}_R \, d\mathbf{r} = \frac{1}{\mu_0} \int \left[ \mathbf{B} \cdot \nabla \mathbf{B} - \nabla \frac{B^2}{2} \right] \cdot \mathbf{e}_R \, d\mathbf{r}$$

$$= -\frac{1}{2\mu_0} \int \left[ \nabla \cdot \left( B_\phi^2 \mathbf{e}_R \right) + \frac{B_\phi^2}{2R} \right] d\mathbf{r} \approx -\frac{1}{2\mu_0} \int \nabla \cdot \left( B_\phi^2 \mathbf{e}_R \right) d\mathbf{r} \quad (B7)$$

$$= -\frac{1}{2\mu_0} \int_{S_{in}} \left( B_\phi^2 \mathbf{n} \cdot \mathbf{e}_R \right) dS$$

The approximate form in the middle equation is a consequence of the thin coil approximation. From simple dimensional analysis, it follows that the neglected term is small by $c_S / R_0$ compared to the term that is maintained.

Using the inner surface parametrization introduced above we see that the radial force reduces to

$$F_R = \frac{\pi R_0^2 B_0^2}{\mu_0} \int_0^{2\pi} \frac{Z_\theta}{R} d\theta \quad (B8)$$

A knowledge of the actual coil shape is needed to evaluate $F_R$. For the simple rectangular coil model introduced in the main text, $F_R$ can be easily evaluated since $R =$ constant on the relevant portions of the surface. We obtain

$$F_R = -\frac{4\pi R_0 B_0^2 (\kappa a + b)}{\mu_0} \frac{\varepsilon_B}{1 - \varepsilon_B^2} \quad (B9)$$

The force $F_R$ is balanced by compression stress due to wedging on the inboard side of the magnet. See Fig. B2. Note that the normal force $F_C$ on each face of the wedged surface is given by

$$F_C = \sigma_C A_C = 2\sigma_C(\kappa a + b)c_S \tag{B10}$$

with $\sigma_C$ the portion of the maximum stress $\sigma_{max}$ balancing the centering force. The component of $F_C$ along $R$ from each face is just $F_C \sin(\Delta\phi/2) \approx F_C \Delta\phi/2$. We now add the $R$ directed stresses from both faces and sum over all $N$ magnets recognizing that by definition $N\Delta\phi = 2\pi$. Force balance between the centering force and the compression stress thus requires $F_R + 2\pi F_C = 0$. Substituting yields an expression for $\sigma_C$,

$$\sigma_C = \frac{B_0^2 R_0}{\mu_0 c_S} \frac{\varepsilon_B}{1-\varepsilon_B^2} \tag{B11}$$

- **Magnet forces: The stress thickness $c_S$**

The quantity $c_S$ is now determined by setting the Tresca stress, $\sigma_T + \sigma_C$, equal to its maximum allowable value, $\sigma_{max}$; that is $\sigma_T + \sigma_C = \sigma_{max}$. The thickness of structural material required to support the magnet thus has the value

$$\frac{c_S}{R_0} = \frac{B_0^2}{4\mu_0 \sigma_{max}}\left[\frac{1}{1-\varepsilon_B}\ln\left(\frac{1+\varepsilon_B}{1-\varepsilon_B}\right) + \frac{4\varepsilon_B}{1-\varepsilon_B^2}\right] \sim 0.2 \tag{B12}$$

which justifies our thin coil approximation. Specifically, the thin coil approximation is valid when $B_0^2/\mu_0\sigma_{max} \ll 1$.

**B.3 Magnet current: the current carrying thickness $c_J$**

The current carrying thickness $c_J$ is easily calculated as follows. The total poloidal current $N_{TF}I_{TF}$ that must flow in the TF magnet system to produce a desired magnetic field on axis $B_0$ is determined from the relation

$$B_0 = \frac{\mu_0 N_{TF} I_{TF}}{2\pi R_0} \tag{B13}$$

Here, $N_{TF}$ is the total number of turns in the TF magnet system. The value of $N_{TF}I_{TF}$ can be written in terms of the maximum allowable current density $J_{max}$ that can safely flow in an HTS tape. Typically $J_{max} \approx 600 - 800$ A/mm$^2$. The relation between $N_{TF}I_{TF}$ and $J_{max}$ is given by

$$\begin{aligned} N_{TF}I_{TF} &= J_{max} A_{HTS} \\ A_{HTS} &= \pi\left[(R_0 - a - b)^2 - (R_0 - a - b - c_J)^2\right] \approx 2\pi R_0(1-\varepsilon_B)c_J \end{aligned} \tag{B14}$$

A simple calculation then leads to

$$\frac{c_J}{R_0} = \frac{B_0}{\mu_0 R_0 J_{max}} \frac{1}{1-\varepsilon_B} \sim 0.01 \tag{B15}$$

Because of the high current carrying capacity of REBCO tapes, the usual situation is that $c_J \ll c_S$. Even so, the tapes are very expensive relative to structural material so it is useful to estimate the number $N_{HTS}$ and total length $L_{HTS}$ required. For the rectangular TF model a straightforward calculation yields

$$\begin{aligned} N_{HTS} &= \frac{2\pi R_0 B_0}{\mu_0 J_{max} hw} \\ L_{HTS} &= \frac{2\pi R_0 B_0}{\mu_0 J_{max}} \frac{[8b + 4(\kappa+1)a]}{hw} \end{aligned} \tag{B16}$$

where $h$ and $w$ are the height and width of the cross section of each tape.

## B.4 The copper and coolant thicknesses $c_{CU}$ and $c_{He}$

The amount of copper $c_{CU}$ to prevent overheating in case of a partial or full quench plus the amount of helium coolant $c_{He}$ to keep the HTS superconducting depends on the details of the specific design under consideration. It is thus difficult to obtain results that are both accurate and general. We avoid this difficulty by examining some earlier HTS studies as well as some LTS studies. This allows us to make simple estimates for both $c_{CU}$ and $c_{He}$. The actual values are not too critical since the overall thickness of the TF coil is dominated by the structural material.

For the equivalent thickness of copper we set

$$c_{CU} \approx 1.6\, c_J \tag{B17}$$

The copper is about 60% thicker than the superconducting tapes. Similarly, for the equivalent coolant thickness we set

$$c_{He} \approx 0.4\, c_{CU} \tag{B18}$$

The coolant thickness is approximately 40% that of the copper.

From these results we see that that the total magnet thickness is given by

$$c = c_S + c_J + c_{CU} + c_{He} \approx c_S + 3\, c_J \tag{B19}$$

and that the ratio

$$\frac{3c_J}{c_S} = \frac{12\sigma_{max}}{\mu_0 R_0 B_0 J_{max}} \frac{1}{\ln\left(\dfrac{1+\varepsilon_B}{1-\varepsilon_B}\right) + \dfrac{4\varepsilon_B}{1+\varepsilon_B}} \approx \frac{2\sigma_{max}}{\mu_0 R_0 B_0 J_{max} \varepsilon_B} \sim 0.1 \tag{B20}$$

As stated earlier, most of the magnet thickness is due to structural support material. Consequently, in order to simplify our design analysis, we shall neglect the $3c_J$ term.

### B.5 Final result

The results discussed above can be combined leading to an expression for the dimensions of a TF magnet.

$$c \approx c_S = \frac{R_0 B_0^2}{4\mu_0 \sigma_{max}} \left[\frac{1}{1-\varepsilon_B} \ln\left(\frac{1+\varepsilon_B}{1-\varepsilon_B}\right) + \frac{4\varepsilon_B}{1-\varepsilon_B^2}\right] \tag{B21}$$

Also, the total height $L_{TF}$ of a TF coil is obviously given by

$$L_{TF} = 2(\kappa a + b + c) \tag{B22}$$

These are the required results.

# Appendix C

# The OH Transformer

The goal of Appendix C is to determine the pulse length and dimensions of the OH transformer as functions of $R_0, B_0$. These dimensions are critical for the design of a pulsed tokamak reactor.

## C.1 Pulse length

The pulse length that must be provided by the OH transformer is determined by three main factors: (1) the number of allowable cycles before replacement is needed, (2) the OH replacement down time, and (3) the need for high average power. The analysis is straightforward and is described below.

In a pulsed tokamak, cyclical thermal and mechanical stresses ultimately cause performance deterioration in the OH transformer. As a result, after $N_{CYC}$ cycles the OH transformer must be replaced. Typically $N_{CYC} \approx 30,000$. Replacement is assumed to take $\tau_{OFF}$ months. During this time, the reactor is off and no power (nor revenue) is being produced. Estimates suggest that $\tau_{OFF} \approx 6$ months. Now, for an economical reactor, the tokamak's operating phase must be much longer than the replacement down time in order to produce high average power over the whole operating-replacement cycle. We denote $N_{ON}$ as the number of down time replacement periods that the reactor must operate to produce high average power; that is, the reactor operating time between OH replacements is $N_{ON}\tau_{OFF}$ months. For an economical reactor we assume $N_{ON} \approx 10$. With these definitions it follows that the number of cycles per month is given by

$$\frac{N_{CYC}}{N_{ON}\tau_{OFF}} \approx 500 \text{ cycles per month} \tag{C1}$$

The pulse length $\tau_P$ is just the inverse of this ratio. Converting from months to hours yields

$$\tau_P = 720\frac{N_{ON}\tau_{OFF}}{N_{CYC}} = 1.44 \sim 1.5 \quad \text{hours per pulse} \tag{C2}$$

with $\tau_{OFF}$ specified in months. We have rounded up the value of $\tau_P = 1.5$ to avoid any false sense of accuracy. The corresponding duty cycle $f_{DC}$ is

$$f_{DC} = \frac{N_{ON}}{N_{ON}+1} \approx 0.9 \tag{C3}$$

Intuitively, we want $\tau_P \propto N_{ON}\tau_{OFF}$ to be as small as possible. Large values require a larger transformer radius to produce the increased demand for transformer volt seconds. This larger radius increases the overall size and capital cost of the reactor.

## C.2 Transformer coil height $L_\Omega$

In sizing the reactor, it is necessary to determine the coil height $L_\Omega$, coil thickness $d$, and inner radius $R_\Omega$ of the OH transformer. A simple estimate for the height $L_\Omega$ is to set it equal to the total vertical dimension of the TF coils. Therefore,

$$L_\Omega = 2(\kappa a + b + c) \tag{C4}$$

## C.3 Transformer coil thickness $d$

The next quantity to be calculated is the thickness of the OH coil $d$ as determined by stress considerations ($d_S$), and to a lesser degree by current carrying capacity ($d_J$), copper protection ($d_{CU}$), and cooling ($d_{He}$). As for the TF coils we write

$$d = d_S + d_J + d_{CU} + d_{He} \tag{C5}$$

Consider first the stresses. There are two main contributions – the tensile stress due to radial expansion forces and compression stress on the top on bottom of the magnet arising from its finite length. The expansion force dominates and for simplicity we neglect the compression forces. Following standard mechanical engineering stress analysis [xx] we find that the local hoop stress in a long cylinder of thickness $d_S$ is given by

$$\sigma_S(R) = \frac{B_\Omega^2}{4\mu_0 \zeta(1 + \zeta/2)}\left[1 + \frac{R_\Omega(1+\zeta)^2}{R^2}\right] \tag{C6}$$

Here, $\zeta = d_S / R_\Omega$ and $B_\Omega$ is the nearly uniform axial magnetic field within the OH transformer.

The area averaged value of $\sigma_S$ must balance the hoop force on the cylinder. Its value, denoted by $\bar{\sigma}_S$, is easily calculated and has the simple form

$$\bar{\sigma}_S = \frac{1}{A}\int \sigma_S dA = \frac{1}{d_S}\int_{R_\Omega}^{R_\Omega + d_S} \sigma_S dR = \frac{B_\Omega^2}{2\mu_0 \zeta} \tag{C7}$$

We now set $\bar{\sigma}_S$ to its maximum allowable value $\hat{\sigma}_{max}$ leading to the desired expression for $d_S$

$$\frac{d_S}{R_\Omega} \equiv \zeta = \frac{B_\Omega^2}{2\mu_0 \hat{\sigma}_{max}} \sim 0.4 \tag{C8}$$

Equation (C8) shows why the thin coil approximation is not accurate for the OH transformer. Also, note that for a pulsed system $\hat{\sigma}_{max} \approx 400 - 500$ MPa as opposed to $\sigma_{max} \approx 600 - 700$ MPa for a steady state system. This is a consequence of the need to extend magnet life due to cyclical stresses.

Next, the magnet thickness required to carry the maximum current that flows in the transformer is easily calculated, as in Appendix B. The basic relation of interest follows from Ampere's law,

$$B_\Omega L_\Omega = \mu_0 N_\Omega I_{max} = \mu_0 \hat{J}_{max} L_\Omega d_J \tag{C9}$$

with $\hat{J}_{max}$ the maximum allowable current density in the OH transformer tapes. It then follows that the thickness $d_J$, number of tapes $N_{HTS}$, and total tape length $L_{HTS}$ are given by

$$\begin{aligned} d_J &= \frac{B_\Omega}{\mu_0 \hat{J}_{max}} \\ N_{HTS} &= \frac{L_\Omega d_J}{wh} = \frac{L_\Omega B_\Omega}{\mu_0 \hat{J}_{max} wh} \\ L_{HTS} &= 2\pi R_\Omega N_\Omega = \frac{2\pi L_\Omega R_\Omega B_\Omega}{\mu_0 \hat{J}_{max} wh} \end{aligned} \tag{C10}$$

Lastly, in analogy with the TF coil we set the copper and cooling thicknesses to $d_{CU} = 1.6\, d_J$ and $d_{He} = 0.4\, d_{CU}$. Therefore, it follows that $d_J + d_{CU} + d_{He} \approx 3 d_J$.

Combining these results, we see that the total thickness of the OH solenoid has the value

$$d = d_S + d_J + d_{CU} + d_{He} \approx d_S + 3 d_J = \frac{R_\Omega B_\Omega^2}{2\mu_0 \hat{\sigma}_{max}} + \frac{3 B_\Omega}{\mu_0 \hat{J}_{max}} \tag{C11}$$

Substituting typical numerical values shows that $3 d_J / d_S \sim 0.2$. As stated, structural requirements dominate current carrying requirements. Thus, without too much loss in accuracy we can neglect the $3 d_J$ term. This, as we shall see, leads to a substantial simplification in the analysis. The end result is that for our design model the thickness of the OH transformer is assumed to be

$$d \approx d_S = \frac{R_\Omega B_\Omega^2}{2\mu_0 \hat{\sigma}_{max}} \tag{C12}$$

**C.4 Transformer inner radius $R_\Omega$**

Finding the radius $R_\Omega$ involves considerable work. The radius is determined by the requirement that the transformer volt-second capacity be sufficiently large to produce a flat-top pulse of desired length $\tau_P$. The transformer also has to provide additional volt-seconds to raise the plasma current from zero to its desired flat-top value $I_M$. In present day short pulse experiments as well as steady state reactors, it is the current rise time requirement that drives the design of the OH transformer.

However, in the multi-hour long pulsed reactors envisaged here, it is plasma maintenance during flat top operation that is the larger, although not dominating, driver of the volt-second requirements. We consequently need to evaluate both the rise time and flat-top volt second requirements since the resulting $R_\Omega$ is a critical quantity determining the major radius $R_0$ of the reactor. For steady state reactors the transformer does not play a major role in setting the value of $R_0$ and, as such, is not discussed any further. The pulsed reactor analysis proceeds as follows.

We apply the integrated form of Faraday's law around a circle whose radius corresponds to the center of the plasma (i.e. $R = R_0$). This leads to

$$\frac{d}{dt}(-\psi_{21} + \psi_{22}) + R_2 I_2 = 0 \tag{C13}$$

Here "1" denotes the $N_\Omega$ turn OH primary and "2" denotes the single turn plasma secondary. Thus, $\psi_{22}$ is the flux contained within $R_0$ due to the plasma current and $\psi_{21}$ is the corresponding flux due to the current flowing in the OH transformer. The negative sign indicates that $\psi_{21}$ is in the opposite direction of $\psi_{22}$. The current $I_2(t) = (1 - f_B)I_P(t)$ represents the Ohmic component of the instantaneous plasma current $I_P(t)$, (i.e. $I_2(t) =$ total plasma current - bootstrap current). Also $\mathrm{R}_2$, using a Roman font, represents resistance in contrast with italicized fonts representing lengths.

We now proceed with a standard electrical engineering analysis where the fluxes are written in terms of the inductances,

$$\begin{aligned}\psi_{22} &= \mathrm{L}_2 I_P \\ \psi_{21} &= \mathrm{M} I_1\end{aligned} \tag{C14}$$

Here $\mathrm{L}_2$ is the plasma inductance and $\mathrm{M}$ is the mutual inductance, again using Roman fonts. For simplicity, both inductances are assumed to be constant in time. Equation (C13) can be rewritten as

$$\mathrm{M}\frac{dI_1}{dt} = \mathrm{L}_2 \frac{dI_P}{dt} + (1 - f_B)\mathrm{R}_2 I_P \tag{C15}$$

To proceed we specify a desired plasma current evolution during the pulse. It consists of a rapid rise from zero to its final value $I$ over a short time scale $\tau_R$. This is followed by a long flat-top period of length $\tau_P$ during which fusion power is being produced. A simple analytic model describing this behavior is given by

$$I_P(t) = (1 - e^{-t/\tau_R})I \qquad 0 < t < \tau_R + \tau_P \tag{C16}$$

To solve for $I_1(t)$ we assume that the primary current must double-swing linearly in time from $-I_{\max}$ to $+I_{\max}$ over the pulse length; that is, $I_1(0) = -I_{\max}$ and $I_1(\tau_R + \tau_P) = +I_{\max}$. Here, $I_{\max}$ is the maximum allowable primary current as set by stress limits on the OH transformer coil. Equation (C15) can now be easily solved for $I_1(t)$

$$I_1(t) = -I_{\max} + \frac{L_2}{M}(1 - e^{-t/\tau_R})I + \frac{(1-f_B)R_2}{M}[t - \tau_R(1 - e^{-t/\tau_R})]I \tag{C17}$$

A sketch of $I_P(t)$ and $I_1(t)$ is illustrated in Fig. C1 for the case where $\tau_R \ll \tau_P$.

The basic transformer relation required for the analysis is now obtained by evaluating Eq. (C17) at $t = \tau_R + \tau_P$ and taking the practical rapid rise time limit $\tau_R \ll \tau_P$. The result is

$$2I_{\max} = \left[L_2 + (1-f_B)R_2\tau_P\right]\frac{I}{M} \tag{C18}$$

The first term represents the rise time flux swing while the second represents the flat-top flux swing.

The next step is to express the quantities in Eq. (C18) in terms of the geometry and plasma properties, which leads to an expression for $R_\Omega$. To begin, we express $I_{\max}$ in terms of $B_\Omega$, the maximum allowable practical magnetic field in the OH transformer,

$$\left. N_\Omega I_1(t)\right|_{\max} = N_\Omega I_{\max} = B_\Omega L_\Omega / \mu_0 \tag{C19}$$

Consider now the flux $\psi_{21} = MI_1$, which can calculated from the Biot-Savart law assuming that the primary current $I_1$ arises from a uniform current density in the CS: $J_\phi(R,Z,t) = N_\Omega I_1(t)/dL_\Omega$. Noting that $\psi_{21} = 2\pi R_0 A_\phi$, we see that the Biot-Savart law reduces to

$$\begin{aligned}
\left.\psi_{21}(R,Z,t)\right|_{R=R_0,Z=0} &= \frac{\mu_0 R_0}{2}\int \frac{J_\phi(R',Z')\cos(\phi'-\phi)}{|\mathbf{r}'-\mathbf{r}|}d\mathbf{r}' \\
&= \frac{\mu_0 R_0 N_\Omega I_1}{2dL_\Omega}\int \frac{\cos(\phi'-\phi)R'dR'dZ'd\phi'}{[(R'+R_0)^2 + Z'^2]^{1/2}[1 - 2k^2\cos(\phi'-\phi)]^{1/2}}
\end{aligned} \tag{C20}$$

where

$$k^2 = \frac{R_\Omega R_0}{[(R_\Omega + R_0)^2 + Z'^2]^{1/2}} \approx \frac{R_\Omega R_0}{[R_0^2 + Z'^2]^{1/2}} \ll 1 \tag{C21}$$

Here, the reasonable, although not great, approximation $R_\Omega \ll R_0$ implies that $k^2$ is small. Exploiting this approximation allows us to evaluate $\psi_{21}$ analytically,

$$\begin{aligned}
\psi_{21}(R,Z,t)\Big|_{R=R_0,Z=0} &\approx \frac{\mu_0 R_0 N_\Omega I_1}{2dL_\Omega} \int \frac{\cos(\phi'-\phi)[1+k^2\cos(\phi'-\phi)]R'dR'dZ'd\phi'}{[(R'+R_0)^2 + Z'^2]^{1/2}} \\
&= \frac{\pi \mu_0 R_0 N_\Omega I_1}{2dL_\Omega} \int_{R_\Omega}^{R_\Omega+d} R'dR' \int_{-L_\Omega/2}^{L_\Omega/2} dZ' \frac{R_0 R'}{[R_0^2 + Z'^2]^{3/2}} \\
&= \frac{\pi \mu_0 N_\Omega I_1}{L_\Omega} \frac{R_\Omega^2}{(1+4R_0^2/L_\Omega^2)^{1/2}} \left(1 + \zeta + \frac{\zeta^2}{3}\right)
\end{aligned} \tag{C22}$$

with $\zeta = d/R_\Omega \approx d_s/R_\Omega$.

The final relation is obtained by setting $\psi_{21} = MI_1$ resulting in an expression for the mutual inductance of the form $M = M(R_\Omega)$

$$M = \frac{\pi \mu_0 N_\Omega R_\Omega^2}{(L_\Omega^2 + 4R_0^2)^{1/2}} \left(1 + \zeta + \frac{\zeta^2}{3}\right) \tag{C23}$$

For the plasma inductance, we shall, for simplicity, use the large aspect circular value first derived by Shafranov. This value is sufficient for present purposes since the flat-top flux swing usually is appreciably larger than the rise time flux swing. Shafranov's result is given by

$$L_2 = \mu_0 R_0 \left[\log\left(\frac{8}{\varepsilon}\right) - 2\right] \tag{C24}$$

Turning to the plasma resistance, we see that a more careful analysis is needed since the current and temperature profiles have smooth, non-uniform, radial dependences over the finite cross section. The resistance can be estimated from the basic power definition

$$R_2 I_2^2 \equiv \int \mathbf{E} \cdot \mathbf{J}_2 d\mathbf{r} \tag{C25}$$

In the large aspect ratio limit, $\mathbf{J} \approx J_2 \mathbf{e}_\phi$ and $\mathbf{E} \approx E_\phi \mathbf{e}_\phi \approx \eta_{NC} J_2 \mathbf{e}_\phi = \eta_{NC}(J_\phi - J_B)\mathbf{e}_\phi$. Here, the neoclassical resistivity in the collisionless limit can be approximated by [zz]

$$\eta_{NC}(\rho) \approx \frac{\eta_S(\rho)}{(1-\varepsilon^{1/2}\rho^{1/2})^2} \qquad \eta_S(\rho) = \frac{3 \times 10^{-8}}{T_k^{3/2}} = \frac{C_\eta}{T_k^{3/2}} \tag{C26}$$

Now, during flat-top operation Faraday's law implies that $E_\phi(\rho) \approx $ constant. Observe, however, that a problem has arisen. Our model profiles for $T_k(\rho)$ and $J_2(\rho)$ do not automatically satisfy the requirement $E_\phi(\rho) = \eta_{NC}(\rho)J_2(\rho) = E_0 = $ constant, particularly near the plasma edge. Rather than defining a whole new set of profiles we can circumvent this problem by assuming that the current density-temperature profiles do actually satisfy the steady state Faraday's law constraint, which will then allow us to determine a direct relation between the plasma resistance and the neoclassical resistivity. In other calculations, the $E_0 = $ constant constraint is not essential since they only involve separate integrals over current density or temperature profiles.

The required resistivity relationship is obtained as follows. The definition of plasma resistance given by Eq. (C25) can be written as

$$R_2 I_2^2 \equiv \int \mathbf{E} \cdot \mathbf{J}_2 d\mathbf{r} \approx 2\pi R_0 \int E_\phi J_2 dA = 2\pi R_0 E_0^2 \int \frac{dA}{\eta_{NC}} = 4\pi^2 R_0 a^2 \kappa E_0^2 \int \frac{\rho\, d\rho}{\eta_{NC}} \tag{C27}$$

The electric field is directly related to the Ohmic current $I_2 = (1-f_B)I$ by

$$I_2 = \int J_2 dA = E_0 \int \frac{dA}{\eta_{NC}} = 2\pi a^2 \kappa E_0 \int \frac{\rho\, d\rho}{\eta_{NC}} \qquad (C28)$$

Eliminating $E_0$ yields

$$R_2 = \frac{R_0}{a^2\kappa} \frac{1}{\int \frac{\rho\, d\rho}{\eta_{NC}}} = \frac{2C_\eta}{\varepsilon^2 \kappa R_0 \overline{T}_k^{3/2}} \frac{1}{G(\varepsilon, \nu_T)}$$

$$G(\varepsilon, \nu_T) = (1+\nu_T)^{3/2} \int_0^1 (1-x)^{3\nu_T/2}(1-\varepsilon^{1/2} x^{1/4})^2\, dx \qquad (C29)$$

$$= (1+\nu_T)^{3/2}\left[\frac{1}{\nu_0} - 2\varepsilon^{1/2}\frac{\Gamma(\nu_0)\Gamma(5/4)}{\Gamma(\nu_0+5/4)} + \varepsilon\frac{\Gamma(\nu_0)\Gamma(3/2)}{\Gamma(\nu_0+3/2)}\right]$$

where $\nu_0 = 1+(3/2)\nu_T$. A plot of $1/G$ versus $\varepsilon$ is plotted in Fig. C2 for $\nu_T = 1.1$. The neoclassical corrections are substantial.

The last quantity of interest is the bootstrap fraction. From the main text, recall that

$$f_B = \frac{I_B}{I} = \frac{2\pi a^2 \kappa}{I}\int_0^1 J_B \rho\, d\rho = K_b \frac{\bar{n}_{20}\overline{T}_k R_0^2}{I_M^2}$$

$$K_b = 0.6099\, \varepsilon^{5/2} \kappa^{1.27}(1+\nu_n)(1+\nu_T)(\nu_n + .054\nu_T) C_B \qquad (C30)$$

The relationships derived above are substituted into Eq. (C18) yielding the desired solution for $R_\Omega$

$$R_\Omega^2 = K_\Omega \left(\frac{L_\Omega^2 + 4R_0^2}{L_\Omega^2}\right)^{1/2} \left(1 - f_B + \frac{\tau_{L/R}}{\tau_P}\right)\left(\frac{1}{1 + \zeta + \zeta^2/3}\right)\frac{I_M}{B_\Omega R_0 \overline{T}_k^{3/2}}$$

$$f_B = K_b \frac{\overline{n}_{20} \overline{T}_k R_0^2}{I_M^2}$$

$$\frac{\tau_{L/R}}{\tau_P} = K_\tau \left[\ln\left(\frac{8}{\varepsilon}\right) - 2\right] R_0^2 \overline{T}_k^{3/2} \tag{C31}$$

$$K_\Omega = 34.38 \frac{\tau_P}{\varepsilon^2 \kappa G}$$

$$K_\tau = 5.818 \times 10^{-3} \frac{\varepsilon^2 \kappa G}{\tau_P}$$

$$\zeta = \frac{d}{R_\Omega} \approx \frac{d_S}{R_\Omega} = \frac{B_\Omega^2}{2\mu_0 \hat{\sigma}_{max}}$$

In this expression $\tau_P$ is given in hours. Typically $R_\Omega \sim 1$ m.

## C.5 Summary of dimensions

Below is a summary of the dimensions of the OH transformer.

**Height**

$$L_\Omega = 2(\kappa a + b + c) \tag{C32}$$

**Thickness**

$$d = d_S + 3d_J \approx d_S$$

$$d_S = \frac{\pi R_\Omega B_\Omega^2}{4\mu_0 \hat{\sigma}_{max}} \tag{C33}$$

$$d_J = \frac{2\hat{\sigma}_{max}}{\hat{J}_{max} B_\Omega R_\Omega} d_S$$

**Radius**

$$R_\Omega^2 = K_\Omega \left(\frac{L_\Omega^2 + 4R_0^2}{L_\Omega^2}\right)^{1/2} \left(1 - f_B + \frac{\tau_{L/R}}{\tau_P}\right)\left(\frac{1}{1+\zeta+\zeta^2/3}\right)\frac{I_M}{B_\Omega R_0 \overline{T}_k^{3/2}} \tag{C34}$$

**C.6 Choice for $B_\Omega$**

The final step in the analysis of the OH transformer is the choice for $B_\Omega$. This is obtained as follows. In terms of the overall size of a pulsed reactor we recall that the constraint to achieve a desired pulse length requires that the transformer radius be sufficiently large to produce the required flux swing. This in turn translates into a requirement on the size of the major radius

$$R_0 = a + b + c + d + R_\Omega \tag{C35}$$

To minimize size and cost we clearly want to minimize $R_0$. Now, in the expression for $R_0$ the only appearance of $B_\Omega$ is in the sum $R_\Omega + d$. In other words, we need to choose $B_\Omega$ to minimize $R_\Omega + d$. Since $B_\Omega \propto \zeta^{1/2}$ this is equivalent to minimizing with respect to $\zeta$. From Eq. (C34) we see that

$$(R_\Omega + d)^2 = R_\Omega^2 (1+\zeta)^2 = \frac{K_\Omega D^2(\zeta)}{(2\mu_0 \hat{\sigma}_{max})^{1/2}} \left(\frac{L_\Omega^2 + 4R_0^2}{L_\Omega^2}\right)^{1/2} \left(1 - f_B + \frac{\tau_{L/R}}{\tau_P}\right)\frac{I_M}{R_0 \overline{T}_k^{3/2}}$$

$$D(\zeta) = \left[\frac{1}{\zeta^{1/2}} \frac{(1+\zeta)^2}{(1+\zeta+\zeta^2/3)}\right]^{1/2} \tag{C36}$$

The function $D(\zeta)$ is a monotonically decreasing function of $\zeta$ as shown in Fig. C3. This implies that $R_\Omega + d$ is minimized by choosing $\zeta$ as large as is technically possible; that is, the most economical choice for $B_\Omega$ is

$$B_\Omega = \hat{B}_{max} \quad \rightarrow \quad \zeta = \zeta_{max} = \frac{\hat{B}_{max}^2}{2\mu_0 \hat{\sigma}_{max}} \tag{C37}$$

For $\hat{B}_{max} = 23$ T and $\hat{\sigma}_{max} = 500$ MPa, then $\zeta = 0.42$.

This completes the analysis of the OH transformer.

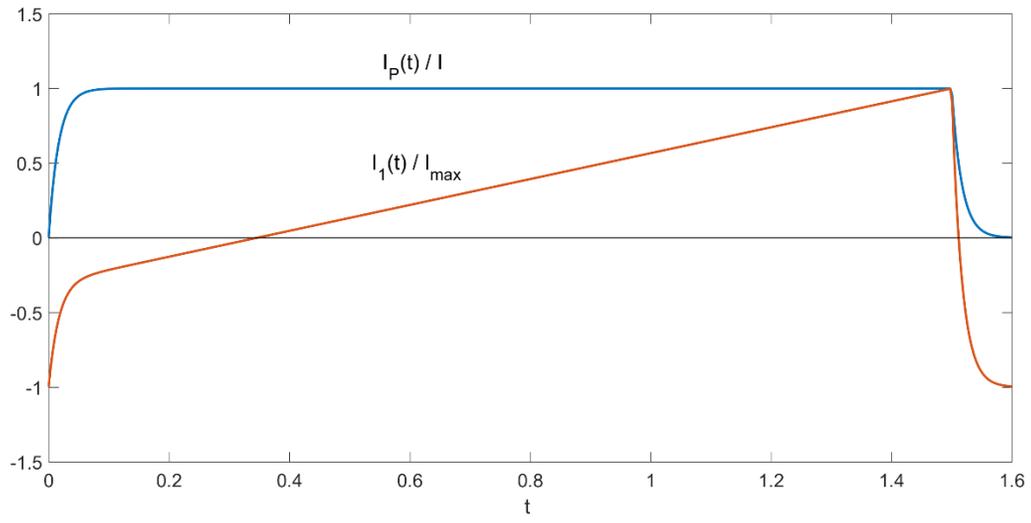

Fig. C1 Sketch of $I_P(t)$ and $I_1(t)$ vs. $t$

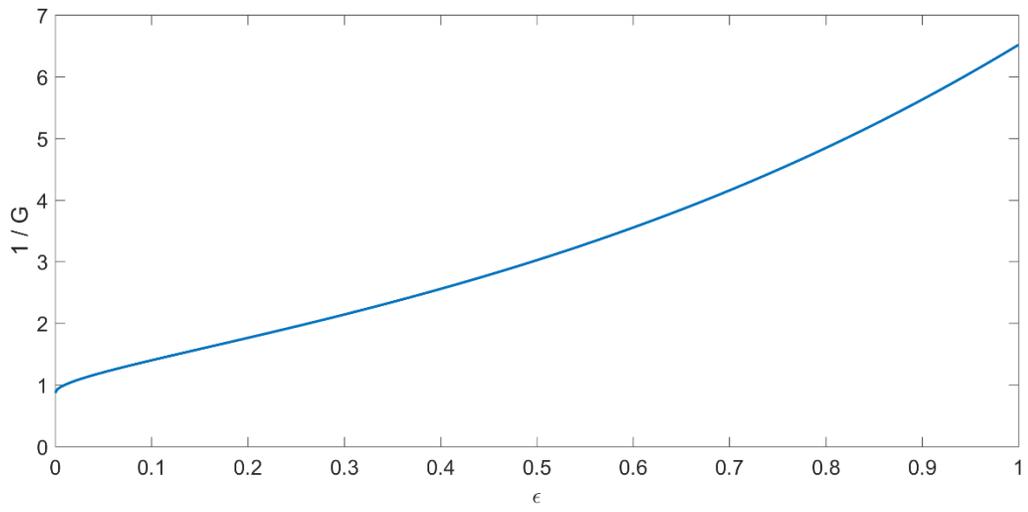

Figure C2 Curve of $1/G$ versus $\varepsilon$ for $\nu_{T} = 1.1$

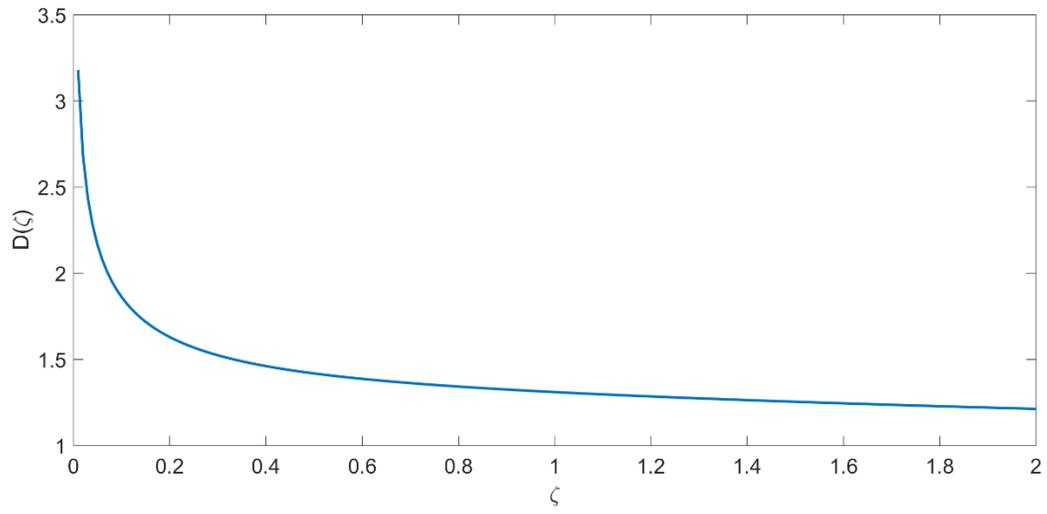

Figure C3 Curve of $D(\zeta)$ versus $\zeta$